\documentclass[11pt]{article}
\usepackage{cite}
\pdfoutput=1
\usepackage{tikz}
\usepackage{pgfplots}
\usepackage{jheppub}
\usepackage{graphicx}
\usepackage{amssymb}
\usepackage{amsmath,amssymb}
\usepackage{slashed}
\usepackage{hyperref}
\usepackage{caption}
\usepackage{xcolor}
\usepackage{dsfont}
\usepackage{verbatim}
\usepackage{subfig}
\usepackage{enumitem}
\usepackage{tikz}

\newcommand\beq{\begin{equation}}
\newcommand\eeq{\end{equation}}
\newcommand\bal{\begin{aligned}}
\newcommand\eal{\end{aligned}}

\def\arsinh{\operatorname{arcsinh}}

\title{Partial Reduction and Cosmology at Defect Brane}

\preprint{\today}

\author[a]{Zhi Wang,}
\author[a]{Zekun Xu,}
\author[a]{Shuyan Zhou,}
\author[a,b,c]{Yang Zhou}
\affiliation[a]{Department of Physics, Fudan University, Shanghai 200433, China}
\affiliation[b]{Center for Field Theory and Particle Physics, Fudan University, Shanghai 200433, China}
\affiliation[c]{Peng Huanwu Center for Fundamental Theory, Hefei, Anhui 230026, China}

\abstract{
Partial reduction is a Randall-Sundrum reduction for only part of the AdS region between finite tension brane and zero tension brane. This is interesting in AdS/BCFT where the AdS bulk contains a defect brane. We employ partial reduction for a AdS bulk with a brane evolving as a $2d$ Friedmann-Robertson-Walker (FRW) cosmology and demonstrate the equivalence between defect extremal surface and island formula for a large subregion fine grained entropy in boundary CFT. We then move to higher dimensions and demonstrate the existence of $4d$ massless graviton on AdS$_4$ brane in partial reduction. We also propose a partial reduction for a $4d$ FRW cosmology at defect brane and obtain the Newton constant by computing boundary entropy.

}

\begin{document}
\maketitle

\section{Introduction}

To find the theory of quantum gravity is one of the most challenging problems in modern physics. In particular, it is neccesary for the microscopic description of black holes as well as our universe. AdS/CFT correspondence provides a promising framework to understand quantum gravity in asymptotically AdS spacetime, which is however different from the universe we observe. It is therefore interesting to ask how to study cosmology in the framework of AdS/CFT. In this paper we initiate a systematical study of Friedmann-Robertson-Walker (FRW) cosmology on the end of the world brane in AdS space using {\it partial reduction.}

We are also motivated by the recent progress of understanding the dynamics of a gravity system glued to a quantum bath, which leads to a theoretical derivation of Page curve for a evaporating black hole~\cite{Penington:2019npb,Almheiri:2019psf}. The evaporating model can be embedded in AdS with the end of the world (EOW) brane, first explored in~\cite{Almheiri:2019hni}, which motivates the so called island formula, later justified in~\cite{Penington:2019kki,Almheiri:2019qdq}. In particular, it was demonstrated~\cite{Rozali:2019day,Deng:2020ent,Chu:2021gdb} in the context of AdS/BCFT that the evaporating black hole can emerge on the Lorentzian EOW brane and the holographic entropy result including the brane matter correction perfectly agrees with island formula. The key step to show the agreement is to employ {\it partial reduction} for AdS space with a brane, which suggests that one should consider the two regions of AdS space bounded by the brane separately: we do partial Randall-Sundrum reduction for the region between zero tension brane and the bounding brane, and employ the ordinary AdS/CFT duality for the remaining region. The decomposition procedure also implies a complete microscopic description of the AdS space with a brane if we dualize everything in the bulk including the brane to the BCFT on the asymptotic boundary. Based on this development, it is then interesting to ask what if we place FRW cosmology on the EOW brane?

In this note we consider a gravitational cosmology on the EOW brane from partial reduction, entangled with a flat CFT on the asymptotic boundary. In Euclidean signature, the model is very similar to static AdS/BCFT~\cite{Takayanagi:2011zk,Fujita:2011fp}, but with spatial and time directions exchanged. Therefore we are actually considering BCFT on a Euclidean time interval. Supposing the Euclidean time is globally compact, the dual bulk will contain a black hole. Solving the brane trajectory and analytically continuing to Lorentzian, one can find that a FRW cosmology emerges on the brane. This model was first proposed by S. Cooper et al. in Ref.~\cite{Cooper:2018cmb}, and the Euclidean brane was later interpreted as bra-ket wormhole in~\cite{Chen:2020tes}. Compare with those works, there are several major differences here: First, we treat the brane as a defect in AdS and include the conformal matter on the brane, therefore the holographic entanglement entropy has to be improved to include the defect contribution. Second, our gravitational action of the Euclidean bra-ket wormhole purely comes from partial reduction, rather than being constructed in the beginning.

We first consider the cosmology from partial reduction in $d=2$. In particular we compute the fine grained entropy for a large subregion in CFT bath from two different approaches: One is the bulk defect extremal surface formula and the other is island formula. We compare those results and find exact agreement. We then move to $d=4$. We show that partial reduction leads to a normalizable zero mode (massless graviton) on Karch-Randall brane in pure AdS$_5$. For the $4d$ FRW cosmology, the brane trajectory is slightly complicated. We propose the proper partial reduction for the $4d$ cosmology. Treating the Euclidean cosmology as the holographic dual for a CFT defect, we compute the boundary entropy for the cosmology and then obtain a Newton constant on the brane. Our partial reduction method provides a new path towards describing our observational cosmology holographically as well as quantum mechanically.

\section{Cosmology from partial reduction in $d=2$}\label{sec2}

In this section, we consider the two-dimensional cosmology at the end of the world brane and compute the entanglement entropy from two different approaches. In the first approach we treat the brane as a defect in the AdS bulk, where a two-dimensional conformal field theory is living on. In the second approach we consider the gravitational cosmology on the brane from partial reduction and use island formula.
We will first review the model in the context of AdS$_{d+1}$/BCFT$_{d}$~\cite{Takayanagi:2011zk,Fujita:2011fp} with particular attention to $d=2$. Then, we recall the defect extremal surface (DES) formula proposed in Ref.~\cite{Deng:2020ent} and use defect extremal surface to calculate the entanglement entropy. In the end we apply the partial reduction procedure to obtain a two-dimensional cosmology in the boundary description, where we will use island formula to calculate the entanglement entropy\footnote{For related works on island of cosmology, see~\cite{Hartman:2020khs,Chen:2020tes,Balasubramanian:2020xqf,Miyaji:2021lcq}.}. We compare the result from DES formula with that from island formula and find exact agreement.

\subsection{Review of cosmology at the end of the world brane}\label{2.1}
\begin{figure}
  \centering
  \includegraphics[scale=0.4]{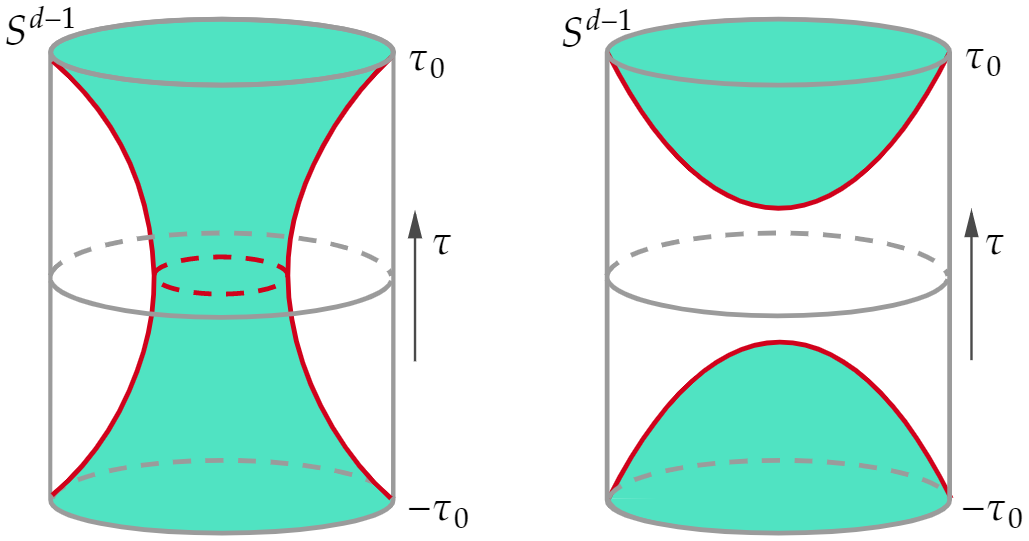}\\
  \caption{Left:~connected EOW brane in Euclidean AdS$_{d+1}$-Schwarzschild geometry. Right:~disconnected EOW brane in thermal-AdS$_{d+1}$.}\label{FIGAdSBCFT}
\end{figure}

Let us review the model we are considering. The model was first explored\footnote{Based on previous works~\cite{Kourkoulou:2017zaj,Almheiri:2018ijj} and further explored in ~\cite{Antonini:2019qkt,VanRaamsdonk:2021qgv,Antonini:2021xar}.} in Ref.~\cite{Cooper:2018cmb}. A boundary conformal field theory (BCFT) is a conformal field theory (CFT) with conformal boundaries (where conformal boundary conditions are imposed)~\cite{Cardy:1989ir,McAvity:1995zd,Cardy:2004hm}. To find the holographic dual for BCFT, Takayanagi proposed a AdS bulk with a constant tension brane, the so called AdS$_{d+1}$/BCFT$_d$ duality~\cite{Takayanagi:2011zk,Fujita:2011fp}. In Euclidean signature, one can consider a holographic BCFT$_d$ defined on an interval of Euclidean time, $[-\tau_0,\tau_0]\times S^{d-1}$. The conformal boundary conditions of the BCFT$_d$ correspond to two boundary state $|B\rangle_{+} $ and $|B\rangle_{-} $, where we use $\pm$ to label the boundary condition at $ \tau=\pm\tau_{0}$. Using AdS$_{d+1}$/BCFT$_d$ duality, this holographic BCFT$_d$ will be dual to an AdS$_{d+1}$ space with a constant tension brane, the so called end of the world (EOW) brane, as shown in Fig.~\ref{FIGAdSBCFT}. The EOW brane can be connected or disconnected, attached to the BCFT$_d$ boundary~\cite{Fujita:2011fp}.
The bulk Euclidean action is
\beq\label{action}
	I_E=-\frac{1}{16\pi G^{(d+1)}_N}\int_N d^{d+1} x\sqrt{g}( R-2\Lambda_{d+1} ) -\frac{1}{8\pi G^{(d+1)}_N}\int_Q d^{d} y\sqrt{h}(K-(d-1)T)\ ,
\eeq
where $N$ represents the bulk and $Q$ represents the EOW brane, we omit the Gibbons-Hawking term on asymptotic boundary. $T$ is the constant tension of the EOW brane and we will see that it is related to the conformal boundary condition. The key assumption for the AdS$_{d+1}$/BCFT$_{d}$ is that the induced metric $h_{ab}$ on the EOW brane preserves the boundary symmetry $SO(d,1)$ and satisfies Neumann boundary condition. The variation of the action to $h_{ab}$ leads to the Neumann boundary condition
\beq\label{NBC}
K_{ab}-Kh_{ab}=(1-d)Th_{ab}\ .
\eeq
Since the holographic BCFT$_d$ defined on $S^{d-1}$ preserves spherical symmetry, the bulk dual will also preserve this symmetry. As a result, it must locally be described by Euclidean AdS$_{d+1}$-Schwarzschild geometry
\beq\label{bme}
	ds^{2} =f( r) d\tau ^{2} +\frac{dr^{2}}{f( r)} +r^{2} d\Omega_{d-1}^{2}\ ,
\eeq
where
\beq
	f(r)=\frac{r^2}{\ell^2}+1-\frac{r_h^{d-2}}{r^{d-2}}(\frac{r_h^2}{\ell^2}+1)\ .
\eeq
$r_h$ is the horizon radius and $\ell$ is the AdS radius. Notice that the Euclidean time circle of black hole is compact and $[-\tau_0,\tau_0]$ should be considered as part of the $\tau$-circle. The brane attached at $\tau=\pm\tau_{0}$ will also preserve the spherical symmetry and we can parameterize it as $r(\tau)$. Using the Neumann boundary condition eq.~(\ref{NBC}), we can find the brane trajectory satisfies
\beq\label{beq}
	\frac{dr}{d\tau}=\pm\frac{f(r)}{Tr}\sqrt{f(r)-T^{2}r^{2}}\ .
\eeq

The bulk metric eq.(\ref{bme}) generally have two different phases. It is known that $r_h=0$ corresponds to thermal-AdS$_{d+1}$ and $r_h>0$ corresponds to AdS$_{d+1}$-Schwarzschild. The transition between the two phases is known as Hawking-Page transition~\cite{Hawking:1982dh,Witten:1998zw}. The Hawking-Page transition will also affect the EOW brane~\cite{Fujita:2011fp,Cooper:2018cmb,Sully:2020pza,Miyaji:2021ktr}. In the thermal AdS$_{d+1}$ bulk, we will have two disconnected EOW branes. While in the AdS$_{d+1}$-Schwarzschild bulk, the EOW brane will be connected.
This is a high temperature phase which dominates for a short time interval $[-\tau_0,\tau_0]$. We will be mostly interested in this phase, shown in left of Fig.~\ref{FIGAdSBCFT}.

Since the Euclidean brane solution of eq.~(\ref{beq}) is symmetric about the $\tau=0$ slice, we can take the $\tau=0$ bulk slice as  initial time slice for a Lorentzian evolution. This is equivalent to apply an analytical continuation $t=-i\tau$ in the time direction. The Lorentzian bulk geometry is thermal-AdS$_{d+1}$ spacetime or AdS$_{d+1}$-Schwarzschild black hole. The brane trajectory satisfies
\beq
	\frac{dr}{dt}=\pm\frac{f(r)}{Tr}\sqrt{T^{2}r^{2}-f(r)}\ .
\eeq
The brane world-volume metric is given by
\beq
    \bal
        ds^2&=-f(r)dt^2+\frac{f(r)(T^{2}r^{2}-f(r))}{T^2r^2}dt^2+r^2d\Omega_{d-1}^2\\
        &=-\frac{f(r)^2}{T^2r^2}dt^2+r^2d\Omega_{d-1}^2\\
        &=-dt_p^2+r^2(t_p)d\Omega_{d-1}^2\ ,
    \eal
\eeq
where in the last step we define a new time $t_p$ satisfying $dt_p=\frac{f(r(t))}{Tr(t)}dt$ and $r(t_p)$ is the scale factor. Below we can find that in the AdS$_{d+1}$-Schwarzschild black hole case, the scale factor corresponds to a big-bang big-crunch cosmology. For general dimensions, the brane trajectory need to be solved numerically. Now we focus on $d=2$ case where everything can be solved analytically. The $d=4$ case will be discussed in next section.

For $d=2$, the boundary state $|B\rangle_{\pm}$ is known as Cardy state~\cite{Cardy:1989ir}. The bulk is an Euclidean BTZ black hole~\cite{Banados:1992wn},
\beq\label{BTZ}
	ds_{\mathrm{EBTZ}}^{2}=\frac{r^2-r_h^2}{\ell^2} d\tau ^{2} +\frac{\ell^2}{r^2-r_h^2}dr^{2} +r^{2} d\theta^{2}\ .
\eeq
The periodicity of the $ \tau $ direction determines the temperature of the black hole, which is
\beq
	\beta =\frac{2\pi \ell^2}{r_h}\ .
\eeq
The brane trajectory $r(\tau)$ satisfies eq.~(\ref{beq}),
\beq
	\frac{dr}{d\tau}=\pm\frac{r^2-r_h^2}{T\ell^2r}\sqrt{\frac{r^2-r_h^2}{\ell^2}-T^2r^2}\ .
\eeq
After integration we obtain the brane trajectory as
\beq\label{BTr}
	\bal
		r( \tau ) & =\frac{r_h}{\sqrt{1-T^{2}\ell^2}}\sqrt{1+T^{2}\ell^2\tan^{2} \frac{r_h \tau }{\ell^2}}\\
		& =\frac{r_h}{\sqrt{1-T^{2}\ell^2}}\frac{1}{\cos \frac{r_h \tau' }{\ell^2}}\ ,
	\eal
\eeq
where we have replaced $ \tau $ with $ \tau'$ by
\beq\label{tautaup}
	\tan \frac{r_h \tau'}{\ell^2}=T\ell\tan \frac{r_h \tau }{\ell^2}\ .
\eeq
We will see that $\tau'$ is the brane conformal time. Notice that $\tau '$ also belongs to $[-\tau _{0} ,\tau _{0} ]$.

\begin{figure}
  \centering
  \includegraphics[scale=0.45]{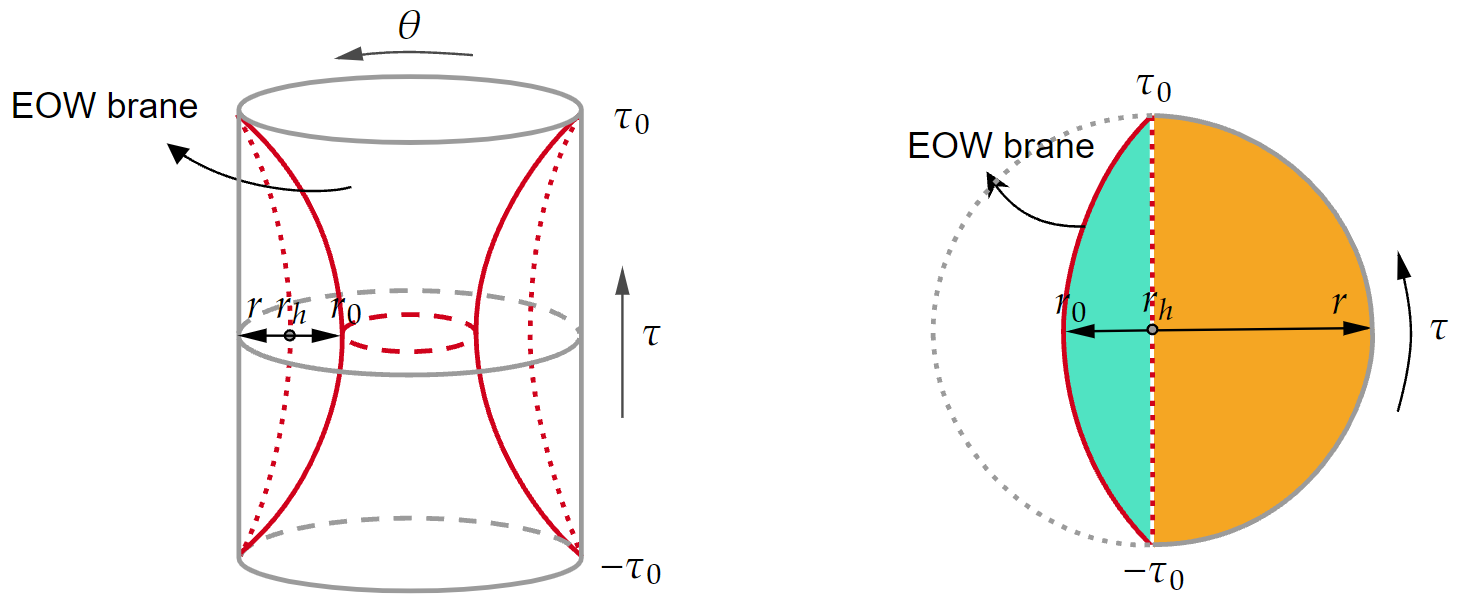}\\
  \caption{Left: Euclidean geometry for the brane trajectory. The red dotted line is the zero tension brane. The red solid line represents any $T>0$ brane. Right: $(\tau,r)$ disc expression, where each point has a $S^1$ of radius $r$.}\label{FIGBranetrajectory}
\end{figure}

It is easy to see that the brane trajectory is symmetric about $\tau=0$, where $r(\tau )$ takes its minimal value, which equals to
\beq
	r_{0} =\frac{r_h}{\sqrt{1-T^{2}\ell^2}}\ .
\eeq
The Euclidean time $\tau _{0}$ can be obtained by subtracting half of the brane time from half of the whole time. It is related to the black hole temperature as
\beq
	\bal
		\tau _{0} & =\frac{\beta }{2} -\int _{r_{0}}^{\infty }\frac{T\ell^2rdr}{(r^{2} -r_h^{2} )\sqrt{\frac{r^2-r_h^2}{\ell^2}-T^2r^2}}\\
		& =\frac{\beta }{2} -\left(\frac{\pi \ell^2}{2r_h} -\frac{\ell^2\arctan\frac{\sqrt{r_{0}^{2} -r{_{H}}^{2} - T^{2}\ell^2r_{0}^{2}}}{ T\ell r_h}}{r_h}\right)\\
		& =\frac{\beta }{2} -\frac{\pi \ell^2 }{2r_h}\\
		& =\frac{\pi \ell^2}{2r_h}\\
		& =\frac{\beta }{4}\ .
	\eal
\eeq
It means that the brane trajectory always meets the boundary at antipodal points in the $(r,\tau)$ disc. The brane trajectory is shown in Fig.~\ref{FIGBranetrajectory}.

Substituting the brane trajectory eq.~(\ref{BTr}) into the bulk metric eq.~(\ref{BTZ}), we obtain the induced metric on the brane as
\beq
	\bal
		ds_{brane}^{2} & =\frac{(r^{2} -r_h^{2} )^{2}}{T^{2}\ell^4r^{2}} d\tau ^{2} +r^{2} d\theta ^{2}\\
		& =\frac{r_h^{2}}{1-T^{2}\ell^2}\frac{T^{2} (1+\tan^{2} \frac{r_h \tau }{\ell^2})^{2}}{1+T^{2}\ell^2\tan^{2} \frac{r_h \tau }{\ell^2}} d\tau ^{2} +\frac{r_h^{2}}{1-T^{2}\ell^2} (1+T^{2}\ell^2\tan^{2} \frac{r_h \tau }{\ell^2})d\theta ^{2}\\
		& =\frac{r_h^{2}}{1-T^{2}\ell^2}\frac{(1+\tan^{2} \frac{r_h \tau }{\ell^2})^{2}}{1+T^{2}\ell^2\tan^{2}\frac{r_h \tau }{\ell^2}}\frac{\cos^{4} \frac{r_h \tau }{\ell^2}}{\cos^{4} \frac{r_h \tau'}{\ell^2}} \frac{1}{\ell^2}d\tau ^{\prime 2} +\frac{r_h^{2}}{1-T^{2}\ell^2} (1+T^{2}\ell^2\tan^{2} \frac{r_h \tau }{\ell^2})d\theta ^{2}\\
		& =\frac{r_{0}^{2}}{\cos^{2} \frac{r_h \tau' }{\ell^2}} (\frac{1}{\ell^2}d\tau ^{\prime 2} +d\theta ^{2} )\ .
	\eal
\eeq
In the third line, we have used $\tan \frac{r_h \tau' }{\ell^2}=T\ell\tan \frac{r_h \tau }{\ell^2}$. And from the last line, we can easily check that it is a hyperbolic space with scalar curvature $R_{brane} =-\frac{2(1-T^2\ell^2)}{\ell^2}$. This metric is conformally equivalent to a finite cylinder metric $ds_{brane}^{2}=\Omega^{-2} (\tau ')ds_{cyl}^{2} =\Omega^{-2} (\tau ')(\frac{1}{\ell^2}d\tau ^{\prime 2} +d\theta ^{2})$ with a conformal factor $\Omega (\tau ')=\frac{1}{r_{0}}\cos \frac{r_h \tau '}{\ell^2}$.

Let us apply an analytical continuation $t=-i\tau$ and the brane trajectory is
\beq\label{LBTr}
	\bal
		r(t) & =\frac{r_h}{\sqrt{1-T^{2}\ell^2}}\sqrt{1-T^{2}\ell^2\tanh^{2} \frac{r_h t }{\ell^2}}\\
		& =\frac{r_h}{\sqrt{1-T^{2}\ell^2}}\frac{1}{\cosh \frac{r_h t' }{\ell^2}}\ ,
	\eal
\eeq
where $t'$ is defined as $\tanh \frac{r_h t'}{\ell^2}=T\ell\tanh \frac{r_h t }{\ell^2}$. Using
\beq	\frac{dr}{dt_p}=\frac{dr}{dt}\cdot\frac{dt}{dt_p}=\pm\sqrt{T^2r^2-\frac{r^2-r_h^2}{\ell^2}}\ ,
\eeq
the scale factor is obtained as $r(t_p)=r_0\cos\frac{r_ht_p}{\ell r_0}$, which describes to a big-bang big-crunch cosmology.

\subsection{Bulk defect extremal surface}

Quantum extremal surface was proposed to compute the holographic entanglement entropy including the contribution of bulk quantum matter~\cite{Faulkner:2013ana,Lewkowycz:2013nqa,Engelhardt:2014gca}
\beq
S_{A} =\min\left\{{\mathrm{ext}\,{S_{gen}}}\right\} =\min\left\{\mathrm{ext_{\Gamma }}\left[\frac{\operatorname{Area}( \Gamma )}{4G_{N}} +S_{\text {bulk}}\right]\right\}\ ,
\eeq
where $\Gamma$ is a co-dimension two surface in AdS bulk which is homologous with subregion $A$.

\begin{figure}[htbp]
  \centering
  \includegraphics[scale=0.5]{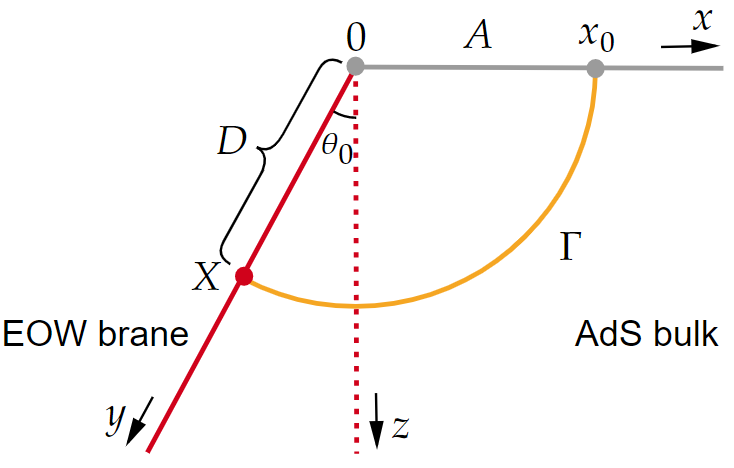}\\
  \caption{The RT surface $\Gamma$ for an interval $A := [0, x_0]$ that contains the boundary. When the boundary of the BCFT is contained in the interval, $\Gamma$ will extend to the EOW brane, on which conformal matter is distributed.}
  \label{FIGDES}
\end{figure}

Quantum extremal surface has been generalized to holographic dualities with defects~\cite{Deng:2020ent}. In quantum field theory with defects, entanglement entropy is generally corrected by the defect. The same thing happens in the bulk. For instance, in~\cite{Takayanagi:2011zk} it was proposed that the entanglement entropy of an interval $[0,x_0]$ on the BCFT can be calculated holographically by the bulk RT surface~\cite{Ryu:2006bv}, as illustrated in Fig.~\ref{FIGDES}. Now if we take the brane as a defect in the bulk and add conformal matter on it, it is obvious that one should include the contribution from these matter when calculating the entanglement entropy. Thus the RT formula should be modified to the defect extremal surface (DES) formula, which is given by \cite{Deng:2020ent}
\begin{equation}\label{DES}
S_A^{\mathrm{DES}}=\min _{\Gamma, X}\left\{\operatorname{ext}_{\Gamma, X}\left[\frac{\operatorname{Area}(\Gamma)}{4 G_{N}}+S_{\text {defect }}[D]\right]\right\}, \quad X=\Gamma \cap D\ ,
\end{equation}
where $\Gamma$ is a co-dimension two surface in AdS bulk and $X$ is the lower dimensional entangling surface determined by the intersection of $\Gamma$ and the defect $D$.

In this note, we will apply the defect extremal surface formula to two-dimensional cosmology at the end of the world brane.

\begin{figure}
	\centering
	\includegraphics[scale=0.5]{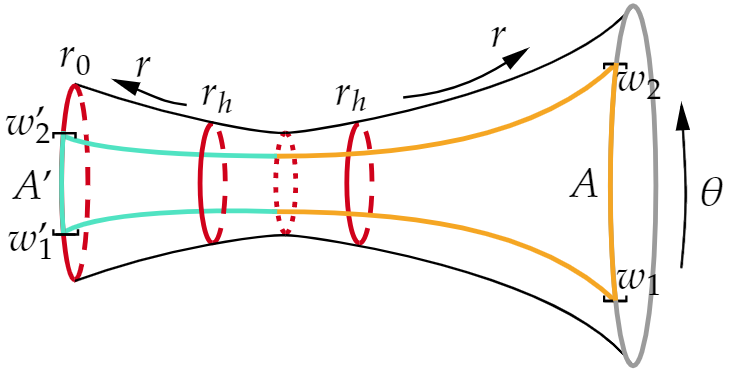}\\
	\caption{An illustration of bulk DES calculation in $t=t_1$ slice.}\label{FIGBulkDES}
\end{figure}

Let us consider the entanglement entropy of a large interval $A$ on the BCFT$_2$.
The end points of the interval $A$ are $w_{1} =(-\theta _{1} ,\frac{\tau _{1}}{\ell} )$ and $w_{2} =(\theta _{1} ,\frac{\tau _{1}}{\ell} )$ at fixed time $\tau=\tau _{1}$ slice. The DES calculation in this time slice is depicted in Fig.~\ref{FIGBulkDES}. Using the defect extremal surface formula, the bulk generalized entropy is

\beq
	S_{gen} =\frac{\text{Area}}{4G_{N}^{(3)}} +S_{A'}\ ,
\eeq
where $S_{A'}$ is the entanglement entropy of the interval in brane conformal field theory.

\begin{figure}
  \centering
  \includegraphics[scale=0.5]{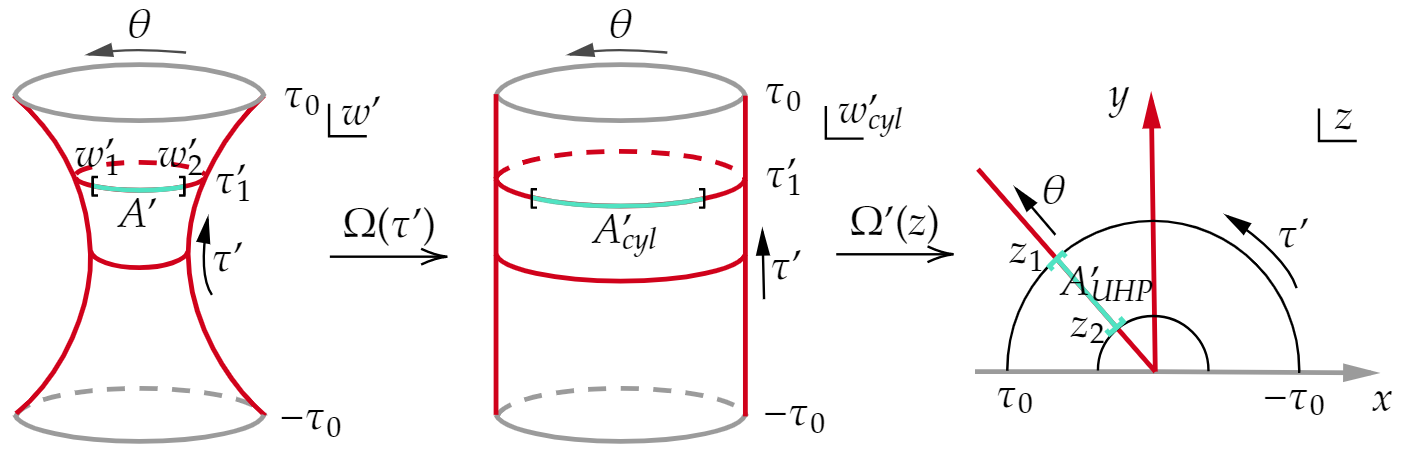}\\
  \caption{Conformal mapping from brane to the UHP. The green line represents the interval.}\label{FIGDESmap}
\end{figure}

In the following, $S_{A'}$ is found to be a constant. Let us use replica trick~\cite{Calabrese:2004eu} to compute $S_{A'}$. The brane metric is
\beq\label{branemetric}
	\bal
		ds_{brane}^{2}=\frac{r_{0}^{2}}{\cos^{2} \frac{r_h \tau '}{\ell^2}} (\frac{1}{\ell^2}d\tau ^{\prime 2} +d\theta ^{2})\ .
	\eal
\eeq
It's conformally equivalent to a cylinder
\beq
	ds_{cylinder}^{2} =\frac{1}{\ell^2}d\tau ^{\prime 2} +d\theta ^{2}\ ,
\eeq
with conformal factor $\Omega ( \tau ') =\frac{1}{r_{0}}\cos \frac{r_h \tau '}{\ell^2}$. The cylinder can be described by complex coordinate $w'=\theta+i\frac{\tau'}{\ell}$. We then map the cylinder onto an upper half-plane (UHP) by
\beq\label{cfm}
	w'=-i\frac{\tau_0}{\ell}+\frac{\ell}{r_h}\log z
\eeq
and the metric becomes
\beq
	ds^{2} =\frac{1}{\ell^2}d\tau ^{\prime 2} +d\theta ^{2} =\frac{\ell^2}{ r_h^{2} z^{2}} dzd\bar{z}\ ,
\eeq
where $z=x+iy$ is the coordinates of the complex plane. The conformal factor is $ \Omega' ( z) =|\frac{r_h z}{\ell}|$. Eventually we map the brane to the UHP and the two boundaries of Euclidean brane now becomes a single boundary. The end points of the interval are mapped to
\beq\label{z1z2}
	z_{1} =e^{-\frac{r_h}{\ell} \theta _{1} +i(\frac{r_h \tau '_{1}}{\ell^2} +\frac{\pi }{2} )}\ ,\;\;\;
	z_{2} =e^{\frac{r_h}{\ell} \theta _{1} +i(\frac{r_h \tau '_{1}}{\ell^2} +\frac{\pi }{2} )}\ .
\eeq
The whole process of the conformal map is illustrated in Fig.~\ref{FIGDESmap}.

\begin{figure}
	\centering
	\includegraphics[scale=0.45]{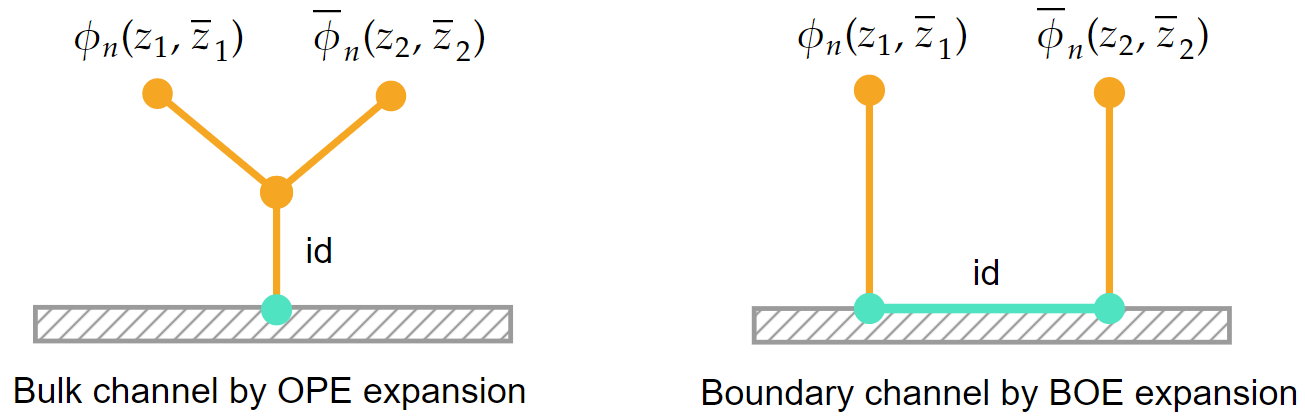}\\
	\caption{The UHP two point function through different channels.}\label{FIGOPEBOE}
\end{figure}

The Renyi entropy for an interval $A'_{\mathrm{UHP}}$ is related to the two point function of twist operators in UHP by
\beq
S_{A'_{\mathrm{UHP}}}^n=\frac{1}{1-n} \log \mathrm{Tr}(\rho_{A'_{\mathrm{UHP}}}^n)=\frac{1}{1-n} \log\langle\phi_n(z_1,\overline{z}_1)\overline{\phi}_n(z_2,\overline{z}_2)\rangle_b.
\eeq
where $b$ represent the conformal boundary condition of the UHP CFT. The doubling trick~\cite{Recknagel:2013uja} indicates that the two point functions in the UHP have the same functional form as the chiral four point functions in the whole plane,
\beq\label{chiral} \langle\phi_n(z_1,\overline{z}_1)\overline{\phi}_n(z_2,\overline{z}_2)\rangle_b
\simeq\langle\phi_n(z_1)\overline{\phi}_n(z_2)\phi_n(\overline{z}_1)\overline{\phi}_n(\overline{z}_2)\rangle.
\eeq
We can expand the two point function in a bulk channel or a boundary channel~\cite{McAvity:1995zd}.
The bulk channel is given by bulk operator expansion (OPE) and the boundary channel is given by boundary operator expansion (BOE). Using doubling trick, the expansion can be expressed as a sum over the conformal blocks of chiral four point function, we refer to Ref.~\cite{Sully:2020pza} for more details.
We sketch these two channels in Fig.~\ref{FIGOPEBOE}. Assuming the vacuum block dominance~\cite{Hartman:2013mia,Faulkner:2013yia,Sully:2020pza}, the two point function in two different channels in the large central charge limit is given by
\begin{align}
\begin{split}
\begin{cases}
    \langle\phi_n(z_1,\overline{z}_1)\overline{\phi}_n(z_2,\overline{z}_2)\rangle_b
    =\frac{g_b^{2(1-n)}|z_1-z_2|^{-2d_n}}{\epsilon'^{-2d_n}},\;\;\;\mathrm{OPE\,\, channel},\\
    \langle\phi_n(z_1,\overline{z}_1)\overline{\phi}_n(z_2,\overline{z}_2)\rangle_b
    =\frac{g_b^{2(1-n)}(4y_1y_2)^{-d_n}}{\epsilon'^{-2d_n}},\;\;\;\mathrm{~BOE\,\, channel},
\end{cases}
\end{split}
\end{align}
where $d_{n} =\frac{c'}{12} (n-\frac{1}{n} )$ is the conformal dimension of the twist operator, $\epsilon '$ is the UV-cutoff and $g_{b}$ is related to the boundary entropy. Combining the conformal factor $(\Omega '(z)\cdot \Omega (\tau '))^{d_{n}} =|\frac{r_h z}{\ell}\cdot \frac{1}{r_{0}}\cos\frac{r_h \tau '}{\ell^2}|^{d_{n}}$, the Renyi entropy of the CFT$_{brane}$ is
\beq\label{Renyi}
    \bal
        S_{A'}^{n} &=\frac{1}{1-n}\log\mathrm{Tr} (\rho _{A'}^{n} )\\
        &=\frac{1}{1-n}\log \bigg[(\Omega'(z_{1} )\cdot \Omega(\tau '_{1} ))^{d_{n}} (\Omega'(z_{2} )\cdot \Omega(\tau '_{1} ))^{d_{n}} \langle \phi _{n} (z_{1} ,\overline{z}_{1} )\overline{\phi }_{n} (z_{2} ,\overline{z}_{2} )\rangle _{b} \bigg]\ .
    \eal
\eeq
The Renyi entropy of $A'$ is obtained by substitute its coordinate eq.~(\ref{z1z2}) into eq.~(\ref{Renyi}).
For the OPE channel, the Renyi entropy is
\beq
	\bal
		S_{A'}^{n} & =\frac{1}{1-n}\log \bigg[(|\frac{r_{0}^{2}}{\cos^{2} \frac{r_h \tau '_{1}}{\ell^2} } \cdot \frac{\ell^2}{r_h^{2} z_{1} z_{2}} |\cdot \frac{|z_{1} -z_{2} |^{2}}{\epsilon ^{\prime 2}} )^{-d_{n}} g_{b}^{2(1-n)} \bigg]\\
		& =\frac{c'}{12}\frac{n+1}{n}\log (\frac{r_{0}^{2}}{\cos^{2} \frac{r_h \tau '_{1}}{\ell^2} } \cdot \frac{\ell^2}{r_h^{2}} \cdot \frac{4\sinh^{2} \frac{r_h \theta _{1}}{\ell} }{\epsilon ^{\prime 2}} )+2\log g_{b}\\
		& =\frac{c'}{6}\frac{n+1}{n}\log\frac{2r_{0}\ell\sinh \frac{r_h \theta _{1}}{\ell} }{r_h\epsilon '\cos \frac{r_h \tau '_{1} }{\ell^2}} +2\log g_{b}\ .
	\eal
\eeq
For the BOE channel, the Renyi entropy is

\beq
	\bal
		S_{A'}^{n} & =\frac{1}{1-n}\log \bigg[(|\frac{r_{0}^{2}}{\cos^{2} \frac{r_h \tau '_{1}}{\ell^2} } \cdot \frac{\ell^2}{r_h^{2} z_{1} z_{2}} |\cdot \frac{4y_{1} y_{2}}{\epsilon ^{\prime 2}} )^{-d_{n}} g_{b}^{2(1-n)} \bigg]\\
		& =\frac{c'}{12}\frac{n+1}{n}\log (\frac{r_{0}^{2}}{\cos^{2} \frac{r_h \tau '_{1}}{\ell^2} } \cdot \frac{\ell^2}{r_h^{2}} \cdot \frac{4\cos^{2} \frac{r_h \tau '_{1}}{\ell^2} }{\epsilon ^{\prime 2}} )+2\log g_{b}\\
		& =\frac{c'}{6}\frac{n+1}{n}\log\frac{2r_{0}\ell}{r_h \epsilon '} +2\log g_{b}\ .
	\eal
\eeq
After taking the limit $n\rightarrow 1$, $S_{A'}$ becomes
\beq
	S_{A'}=\lim _{n\rightarrow 1} S_{A'}^{n} =
    \begin{cases}
		\frac{c'}{3}\log\frac{2r_{0}\ell\sinh \frac{r_h \theta _{1}}{\ell}  }{r_h\cos \frac{r_h \tau '_{1}}{\ell^2} \epsilon '}+2\log g_{b} \ \ \ ,\ \ \ \mathrm{OPE\ \ channel}\ ,\\
        \frac{c'}{3}\log\frac{2r_{0}\ell}{r_h \epsilon '} +2\log g_{b} \ \ \ \ \ \ \ \ \ \ \ \ \ ,\ \ \ \mathrm{BOE\ \ channel}\ .
	\end{cases}
\eeq

For a large enough interval $A$, the entanglement wedge will include the brane region and the BOE channel will be of our interest. In this case, the entropy of the interval is
\beq
	S_{A'} =\frac{c'}{3}\log\frac{2r_{0}\ell}{r_h \epsilon '} +2\log g_{b}\ ,
\eeq
which is a constant.
\begin{figure}
  \centering
  \includegraphics[scale=0.45]{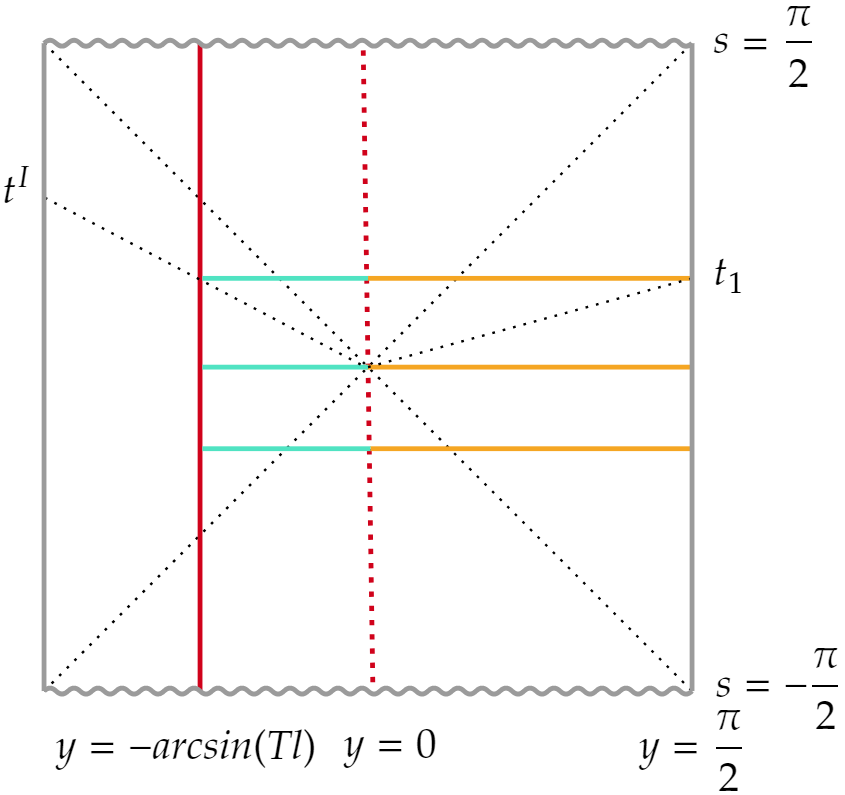}\\
  \caption{BTZ black hole in $s$-$y$ coordinate. The red solid line is the brane with tension $T$, the red dashed line is the brane with tension zero. The line orthogonal to the brane is the RT surface. }\label{FIGRT2D}
\end{figure}
Below we will assume that the central charge $c'$ of the brane CFT is the same as the central charge $c$ of holographic BCFT$_2$ and we take the boundary entropy $\log g_b$ of the brane CFT to be zero, then
\beq
	S_{A'} =\frac{c}{3}\log\frac{2r_{0}\ell}{r_h \epsilon '}\ .
\eeq

To compute $S_A^{\mathrm{RT}}$, let us work in Lorentzian signature and change from Schwarzschild $t$-$r$ coordinate to Kruskal-type $s$-$y$ coordinate by replacing
\beq
	t=\frac{\ell^2}{2r_h}\ln\left( -\frac{\tan\left(\frac{s+y}{2}\right)}{\tan\left(\frac{s-y}{2}\right)}\right), \ \ \ r=r_h\frac{1-\tan\left(\frac{s+y}{2}\right)\tan\left(\frac{s-y}{2}\right)}{1+\tan\left(\frac{s+y}{2}\right)\tan\left(\frac{s-y}{2}\right)}\ .
\eeq
In $s$-$y$ coordinate, the maximally extended BTZ black hole takes the form
\beq\label{syc}
	ds_{\mathrm{BTZ}}^{2} =\frac{1}{\cos^{2}y}( -\ell^2ds^{2} +\ell^2dy^{2} +r_h^{2}\cos^{2}(s) d\theta ^{2})\ .
\eeq
After this transformation, the brane trajectory in $s$-$y$ coordinate is
\beq
	y=-\arcsin(T\ell)\ ,
\eeq
which says that the brane stays in constant $y$ slice shown in red line in Fig. \ref{FIGRT2D}.

The RT surface and the brane are perpendicular to each other at the intersection point. The cutoff surface is $r_{max}=\frac{\ell^2}{\epsilon}$. We then obtain $S_A^{\mathrm{RT}}$ as
\begin{align}
\begin{split}
S_A^{\mathrm{RT}}=\begin{cases}
    \frac{c}{3}\log\frac{2\ell^2\sinh\frac{r_h\theta_1}{\ell}}{r_h\epsilon}\;\;\; ,\;\;\;\mathrm{conn.}\\
    \frac{c}{3}\log\frac{2\ell^2\cosh\frac{r_ht_1}{\ell^2}}{r_h\epsilon}+\frac{c}{3}\log\sqrt{\frac{1+T\ell}{1-T\ell}}\;\;\;,\;\;\;\mathrm{disc.}
\end{cases}
\end{split}
\end{align}
In our case, the disconnected phase dominates for a large interval.

Adding the two terms together, the holographic entanglement entropy calculated by DES formula is
\beq\label{SDES}
S_A^{\mathrm{DES}}=S_A^{\mathrm{RT}}+S_{A'}=\frac{c}{3} \log \frac{2r_0\ell}{r_h\epsilon'}+\frac{c}{3}\log\frac{2\ell^2\cosh\frac{r_ht_1}{\ell^2}}{r_h\epsilon}
+\frac{c}{3}\log\sqrt{\frac{1+T\ell}{1-T\ell}}\ .
\eeq

\subsection{Island formula for cosmology at the end of the world}\label{sec2.3}

Let us justify the entanglement entropy result obtained in the previous subsection by island formula~\cite{Almheiri:2019hni}. An effective $2d$ description for the bulk with a brane can be obtained by combining {\it partial Randall-Sundrum reduction} and {\it AdS/CFT correspondence.} The procedure of partial reduction is as follows. The bulk is decomposed into two parts by the zero tension brane, whose trajectory is
\beq
r(\tau')=\frac{r_h}{\cos\frac{r_h \tau'}{\ell^2}}\ ,
\eeq
which intersects with the bifurcation point at $\tau'=0$. Since no physical degrees of freedom is living on the brane, it can be viewed as a transparent boundary condition.

\begin{figure}
  \centering
  \includegraphics[scale=0.6]{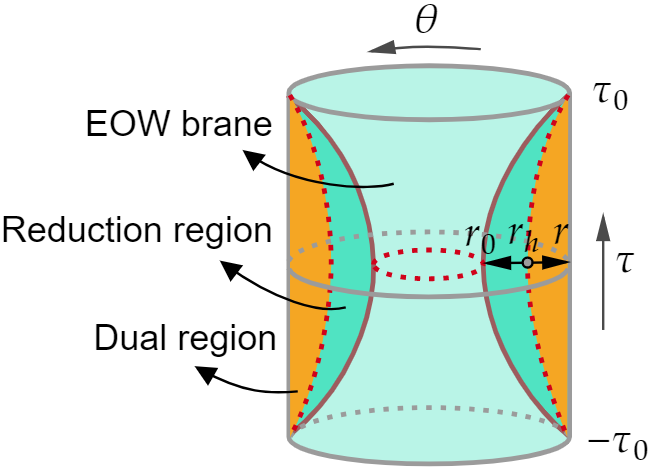}\\
  \caption{The bulk region is divided into two parts by the zero tension brane. A $2d$ effective theory is obtained by applying partial Randall-Sundrum and AdS/CFT to reduction region and dual region respectively.}\label{FIGPartialreduction}
\end{figure}

We divide the bulk region bounded by the finite tension brane into two parts: {\it reduction region} and {\it duality region} as shown in  Fig.~\ref{FIGPartialreduction}. The whole process of partial reduction is as follows. For the reduction region, we employ Randall-Sundrum reduction~\cite{Randall:1999ee,Randall:1999vf} and obtain a gravity theory on the brane, coupled with CFT matter. For the duality region, we use the ordinary AdS/CFT and obtain a boundary conformal field theory with zero boundary entropy. The brane CFT and holographic BCFT$_2$ are glued together with transparent boundary condition, which is essentially the dual of the imaginary bulk zero tension brane.
By integrating the warp factor from the zero tension brane to the $T$ tension brane~\cite{Akal:2020wfl}, the Newton constant on the brane is
\beq
	\bal
		\frac{1}{4G_{brane}} &=\frac{\ell}{4G_{N}^{(3)}}\int _{-\arcsin (T\ell)}^{-\arcsin 0}\frac{dy}{\cos y}\\
		&=\frac{\ell}{4G_{N}^{(3)}} \left[-\log(\cos\frac{y}{2} -\sin\frac{y}{2}) +\log(\cos\frac{y}{2} +\sin\frac{y}{2})\right]\Big |_{-\arcsin (T\ell)}^{-\arcsin 0} \\
		&=\frac{\ell}{4G_{N}^{(3)}}\log\sqrt{\frac{1+T\ell}{1-T\ell}}\ .
	\eal
\eeq

\begin{figure}
  \centering
  \includegraphics[scale=0.5]{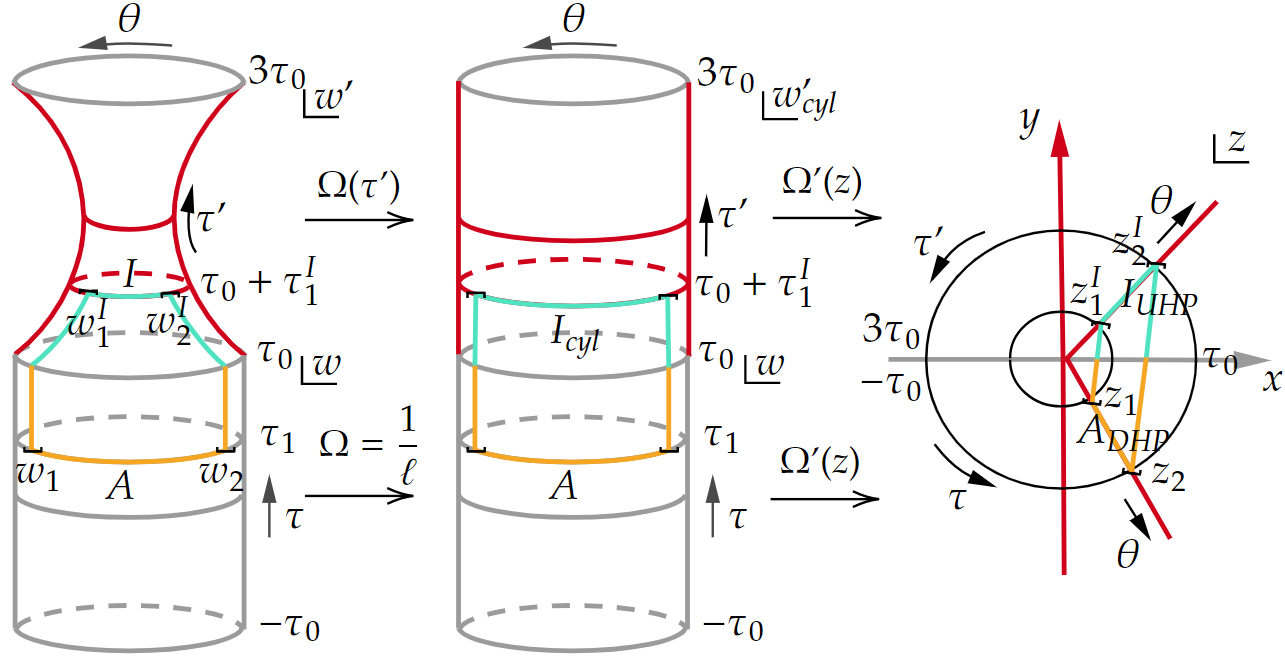}\\
  \caption{Conformal maps from the brane and bulk to a whole complex plane for island calculation.}\label{FIGQESmap}
\end{figure}

Now we can use the island formula to compute the entanglement entropy for the same boundary interval $A$. The generalized entropy is expressed as
\beq
	S_{gen} =\frac{\text{Area}(\partial I)}{4G_{brane}} +S(A\cup I)\ ,
\eeq
where $I$ represents the island region.
We can also view the $2d$ brane after partial reduction as a bra-ket wormhole~\cite{Chen:2020tes}. Notice that in Euclidean signature, the brane time $\tau'$ can be treated as a natural extension of the boundary time $\tau$, as shown in Fig.~\ref{FIGQESmap}. Therefore the island lies between $w_{1}^{I} =(-\theta _{1} ,\frac{\tau _{0} +\tau _{1}^{I}}{\ell} )$ and $w_{2}^{I} =(\theta _{1} ,\frac{\tau _{0} +\tau _{1}^{I}}{\ell} )$.\footnote{For symmetry reason, we will only consider the extremization over Euclidean time direction.} Since the boundary of the island has two points, the area term is
\beq
	\bal
	S' & =\frac{\text{Area}(\partial I)}{4G_{brane}}\\
	& =2\cdot \frac{1}{4G_{brane}}\\
	& =\frac{\ell}{2G_{N}^{(3)}}\log\sqrt{\frac{1+T\ell}{1-T\ell}}\\
	& =\frac{c}{3}\log\sqrt{\frac{1+T\ell}{1-T\ell}}\ ,
	\eal
\eeq
where we have substituted $G_{N}^{(3)}=\frac{3\ell}{2c}$  and taken the area of a point as $1$. It is easy to see that $S'$ is a constant, which means that we only need to extremize $S(A\cup I)$ to determine the location of the island.

In the large interval limit, the entropy $S(A\cup I)$ can be decomposed as
\beq
	S(A\cup I)=S[w_{1} ,w_{1}^{I} ]+S[w_{2} ,w_{2}^{I} ]\ .
\eeq
Similar to the previous section, it can be calculated by twist operator correlation function. Using the conformal transformation in eq.~(\ref{branemetric})(\ref{cfm}), we can map CFT$_{brane}$ to UHP and $w_{1}^{I}$,  $w_{2}^{I}$ to
\beq
	z_{1}^{I} =e^{-\frac{r_h \theta _{1}}{\ell} +i\frac{r_h \tau _{1}^{I}}{\ell^2}}\ ,\;\;\;
	z_{2}^{I} =e^{\frac{r_h \theta _{1}}{\ell} +i\frac{r_h \tau _{1}^{I}}{\ell^2}}\ .
\eeq
The conformal factor is $ \Omega' ( z) \cdot \Omega ( \tau ') =| \frac{r_h z}{\ell}\cdot \frac{1}{r_{0}}\cos \frac{r_h \tau '}{\ell^2}|$. Simultaneously, using the conformal transformation
\beq
	w=i\frac{\tau_0}{\ell}+\frac{\ell}{r_h}\log z\ ,
\eeq
we map the holographic BCFT$_2$ to the down half-plane (DHP) with conformal factor $ \Omega\cdot\Omega '( z) =\frac{1}{\ell}\cdot|\frac{r_h z}{\ell}|$. $w_{1}$ and $w_{2}$ are mapped to $z_{1} =e^{-\frac{r_h \theta _{1}}{\ell} +i(\frac{r_h \tau _{1}}{\ell^2} -\frac{\pi }{2} )}$ and $z_{2} =e^{\frac{r_h \theta _{1}}{\ell} +i(\frac{r_h \tau _{1}}{\ell^2} -\frac{\pi }{2} )}$ as shown in Fig.~\ref{FIGQESmap}.

The entropy between $z_1$ and $z_1^I$ is
\beq
	S[z_1 ,z_1^I] = \frac{c}{6}\log\frac{|z_1^I-z_1|^2}{\epsilon'\epsilon}\ .
\eeq
Combining with the conformal factor, the entropy in the original coordinate is
\beq
	\bal S[w_1,w^I_1]&=\frac{c}{6}\log\frac{|z_1^I-z_1|^2}{\Omega'(z_1^I)\Omega(\tau_1^I)\epsilon'\cdot\Omega'(z_1)\epsilon}\\
            &=\frac{c}{6}\log\frac{|z_1^I-z_1|^2}{| \frac{r_h z_1^I}{\ell}\cdot \frac{1}{r_{0}}\cos (\frac{r_h \tau_1^I}{\ell^2}+\frac{\pi}{2})|\epsilon'\cdot|\frac{r_h z_1}{\ell^2}|\epsilon}\\
            &=\frac{c}{6}\log\frac{r_0\ell^3|e^{i\frac{r_h \tau _{1}^{I}}{\ell^2}}-e^{i(\frac{r_h \tau _{1}}{\ell^2} -\frac{\pi }{2} )}|^2}{\epsilon'\epsilon r_h^2\sin \frac{r_h \tau_1^I}{\ell^2}}\\
            &=\frac{c}{6}\log\frac{2r_0\ell^3(1+\sin\frac{r_h (\tau _{1}^{I}-\tau_1)}{\ell^2})}{\epsilon'\epsilon r_h^2\sin \frac{r_h \tau_1^I}{\ell^2}}\ .
	\eal
\eeq
Since there is no $\theta $ dependence, $ S[w_{1} ,w_{1}^{I} ]$ and  $S[w_{2} ,w_{2}^{I} ]$ are the same. $S(A\cup I)$ is given by
\beq\label{qes}
	S(A\cup I)= S[w_{1} ,w_{1}^{I} ]+S[w_{2} ,w_{2}^{I} ]=2S[w_{1} ,w_{1}^{I} ]\ .
\eeq
Extremizing it with respect to $\tau _{1}^{I}$, the location of the island satisfies
\beq
	\partial _{\tau _{1}^{I}} S(A\cup I) =\frac{\cos\frac{r_h \tau _{1}^{I}}{\ell^2}-\sin\frac{r_h \tau _{1}}{\ell^2}}{(\sin\frac{r_h (\tau_1-\tau _{1}^{I})}{\ell^2}-1)\sin \frac{r_h \tau _{1}^{I}}{\ell^2}}=0\ .
\eeq
The solution is
\beq\label{tau1I}
	\frac{r_h \tau _{1}^{I}}{\ell^2} =\frac{\pi }{2} -\frac{r_h \tau_1}{\ell^2}\ .
\eeq
Back to the brane time, one can check that the island appears in the future to boundary subregion in $\tau$ coordinate. This is in agreement with bulk RT surface shown in Fig.~\ref{FIGRT2D}.

Substituting eq.~(\ref{tau1I}) into eq.~(\ref{qes}), $S(A\cup I)$ takes its extremal value
\beq
	 S_{min}(A\cup I)=\frac{c}{3} \log \frac{2r_0\ell}{r_h\epsilon'}+\frac{c}{3}\log\frac{2\ell^2\cos\frac{r_h\tau_1}{\ell^2}}{r_h\epsilon}\ .
\eeq
So the entanglement entropy for subregion $A$ calculated by island formula is
\beq
	\bal
		S_{A} & =\mathrm{Min}_{I} \ S_{gen}\\
		& =\frac{\mathrm{Area}(\partial I)}{4G_{brane}} +S_{min} (A\cup I)\\
		& =\frac{c}{3} \log \frac{2r_0\ell}{r_h\epsilon'}+\frac{c}{3}\log\frac{2\ell^2\cos\frac{r_h\tau_1}{\ell^2}}{r_h\epsilon} +\frac{c}{3}\log\sqrt{\frac{1+T\ell}{1-T\ell}}\ .
	\eal
\eeq
After the analytic continuation to real time
\beq
S_{A}=\frac{c}{3} \log \frac{2r_0\ell}{r_h\epsilon'}+\frac{c}{3}\log\frac{2\ell^2\cosh\frac{r_ht_1}{\ell^2}}{r_h\epsilon} +\frac{c}{3}\log\sqrt{\frac{1+T\ell}{1-T\ell}}\ ,
\eeq
it matches the DES result eq.~(\ref{SDES}) precisely.

\section{Cosmology from partial reduction in $d=4$}

In this section, we generalize the partial reduction to $4d$ cosmology. We will first study the partial reduction for Karch-Randall brane~\cite{Karch:2000ct} in pure AdS$_5$ space. Based on this, we then propose a proper partial reduction for a $4d$ cosmology at the end of the world brane. We also calculate the boundary entropy associated to the boundary state in the Euclidean cosmology model. Demanding the boundary entropy as an area entropy on the brane after partial reduction, we deduce the Newton constant.

\subsection{Partial Randall-Sundrum reduction in AdS$_5$}

In Randall-Sundrum II reduction~\cite{Randall:1999vf}, we consider a Minkowski brane in AdS$_5$ and the brane tension has to be fine tuned. In particular, there is a massless graviton localized on the brane. However, if we consider AdS$_4$ Karch-Randall brane embedded in AdS$_5$~\cite{Karch:2000ct}, the massless mode is missing because it is non-normalizable. Apart from the brane, the asymptotical boundary of AdS$_5$ in this case also attracts massless graviton significantly. We propose {\it partial reduction} to avoid this problem. In partial reduction, we split the whole bulk into two parts: one is the region between Karch-Randall brane and zero tension brane and the other is the remaining part. The brane gravity only comes from the reduction of the first part and the dynamics of the remaining part can be translated into a boundary conformal field theory coupled to the brane gravity. Note that this does not mean that the full reduction is wrong, rather it means that we can decompose the whole bulk and leave some of the bulk dynamics to the asymptotical boundary. This decomposition procedure has been justified in great detail~\cite{Deng:2020ent,Chu:2021gdb,Li:2021dmf} in the recent understanding of black hole information problem. In particular, it reproduces results from island formula precisely. We will demonstrate here the existence of a normalizable zero mode as well as a well-defined Newton constant in partial reduction. We will also discuss the relation between boundary entropy and Newton constant.

\subsubsection{Localized gravity by partial reduction}

\begin{figure}
	\centering
	\includegraphics[scale=0.5]{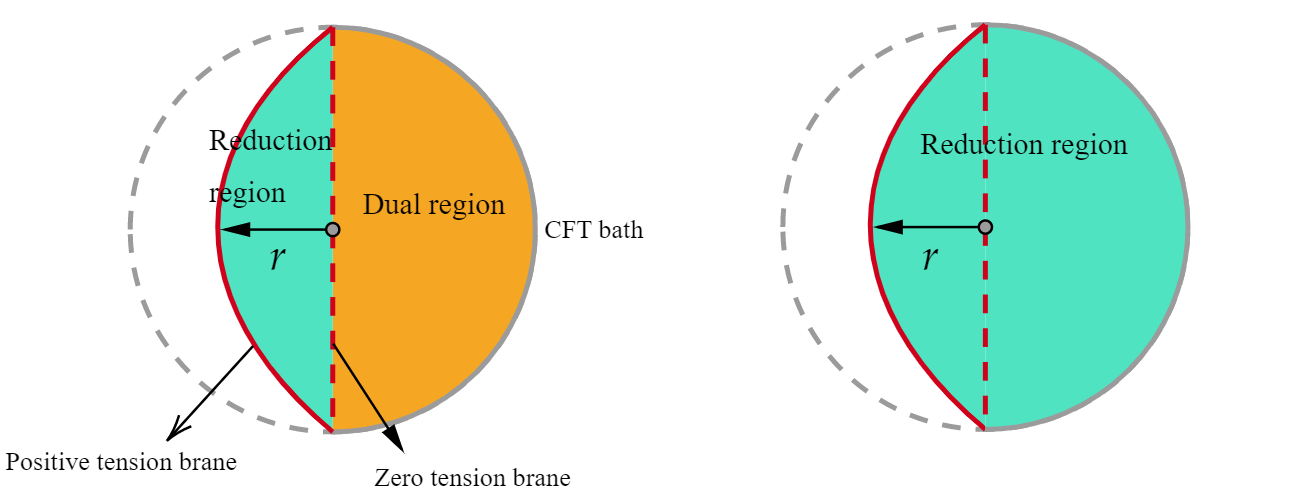}\\
	\caption{The illustration of partial reduction(left) and full reduction(right) of a time slice of AdS$_5$. The red dashed line represents the zero tension brane and red solid line is the positive tension brane.}\label{reduction}
\end{figure}

Let us consider the bulk of AdS$_{d+1}$/BCFT$_d$. The bulk action is
\beq
	I=\frac{1}{16\pi G^{(d+1)}_N}\int_N d^{d+1} x\sqrt{-g}( R-2\Lambda_{d+1} ) +\frac{1}{8\pi G^{(d+1)}_N}\int_Q d^{d} y\sqrt{-h}(K-(d-1)T)\ ,
\eeq
where $N$ represents the bulk and $Q$ represents the EOW brane. $T$ is the constant tension of the EOW brane. The variation of the action with respect to the brane metric $h_{ab}$ leads to the Neumann boundary condition
\beq
K_{ab}-Kh_{ab}=(1-d)Th_{ab}\ .
\eeq
One can easily check that a AdS$_{d+1}$ spacetime with a AdS$_d$ embedding brane is a solution.  For $d=4$, the bulk metric is given by
\beq\label{ads51}
ds^2=dr^2+\cosh^2\frac{r}{\ell}\cdot\frac{\ell^2}{z^2}(-dt ^{2}+dz^2+dx_{1}^{2}+dx_{2}^{2})\ ,
\eeq
with $\ell$ the AdS$_5$ radius. The brane is located at $r=r_*$ with a AdS$_4$ metric, where the location depends on the tension by $T\ell=\tanh\frac{r_*}{\ell}$.
We can do reduction along $r$ direction from $0$ to $r_*$ and obtain a $d$-dimensional gravity on the brane, as shown in Fig.~\ref{reduction}.
The brane effective action includes~\cite{Deng:2020ent}
\beq
\bal
I_{eff}
&\supset\frac{1}{16 \pi G_{N}^{(d+1)}}\left(\cosh \frac{r_*}{\ell}\right)^{2-d} \int_{0}^{r_*} d r\left(\cosh \frac{r}{\ell}\right)^{d-2} \int_{\mathrm{Q}} \sqrt{-g^{(d)}} R^{(d)}\\
&\equiv \frac{1}{16 \pi G_{N}^{(d)}} \int_{\mathrm{Q}} \sqrt{-g^{(d)} }R^{(d)}\ ,
\eal
\eeq
where the superscript $(d)$ denotes the quantities for the effective $d$-dimensional gravity theory on the brane.
One can therefore find the effective $d$-dimensional Newton constant to be
\beq
\frac{1}{G_{N}^{(d)}}=\frac{1}{G_{N}^{(d+1)}}\left(\cosh \frac{r_*}{\ell}\right)^{2-d} \int_{0}^{r_*} d r\left(\cosh \frac{r}{\ell}\right)^{d-2}\ .
\eeq
In $d=4$, this gives
\beq\label{new}
\bal
\frac{1}{G^{(4)}_N}&=\frac{1}{G^{(5)}_N}\frac{1}{\cosh^2(r_*/\ell)}\int _{0}^{r_{*}} \cosh^2(r/\ell)dr\\
&=\frac{1}{G^{(5)}_N}\frac{\left(r_*+\ell \sinh(r_*/\ell) \cosh (r_*/\ell)\right)}{2\cosh^2(r_*/\ell)}\ .
\eal
\eeq
The effective Newton constant $G^{(4)}_N$ depends on the reduction region. In full reduction from $-\infty$ to $r_*$, the integration is divergent, which makes $G^{(4)}_N$ goes to $0$. But in partial reduction, the reduction region is from $0$ to $r_*$, so we obtain a finite Newton constant.

One can also discuss the partial reduction in the set up of Karch-Randall~\cite{Karch:2000ct}. There the brane was introduced as a delta source in the action,
\beq
S=\frac{1}{16\pi G_{N}^{(5)}}\left[\int d^5x\sqrt{|\det g_{\mu\nu}|}(R-2\Lambda_5)+12T\int d^4xdr  \sqrt{|\det g_{ij}|} \delta(r-r_*)\right]\ .
\eeq
With the coordinate transformation
\beq
z=\frac{1}{\sqrt{-\Lambda}}e^{\sqrt{-\Lambda}x_3}\ ,
\eeq
the $5d$ metric eq.~(\ref{ads51}) becomes
\beq
ds^2=dr^2+\cosh^2\frac{r}{ \ell}\cdot(-\Lambda)\ell^2\left[e^{-2\sqrt{-\Lambda}x_3}(-dt ^{2}+dx_{1}^{2}+dx_{2}^{2})+dx_3^2\right]\ ,
\eeq
where the metric in $[\,...\,]$ is a AdS$_4$ metric. One can check that this is indeed the solution in $r\in(0,r_*)$ for the above action with the following ansatz
\beq
ds^2=dr^2+e^{2A(r)}\left[g_{ji}dx^idx^j\right]=dr^2+e^{2A(r)}\left[e^{-2\sqrt{-\Lambda}x_3}(-dt ^{2}+dx_{1}^{2}+dx_{2}^{2})+dx_3^2\right]\ .
\eeq
The equations of motion (EOM) are
\beq
A'^{2} -\Lambda e^{-2A} =\frac{1}{\ell^{2}}\ ,
\eeq
\beq
A''+\Lambda e^{-2A} =2T \delta ( r-r_*)\ .
\eeq

Now we calculate the perturbation of the metric to determine the spectrum following~\cite{Mannheim2005Brane}. Here we consider the transverse-traceless (TT) mode of the gravity fluctuations $\gamma_{ij}$ with $\nabla^i\gamma_{ij}, \gamma_i^i=0$ and impose axial gauge $\gamma_{ir} =0$. The perturbed metric of the AdS$_4$ is
\beq
ds^{2} =dr^2+\left(e^{2A} g_{ij} +\gamma_{ij}\right) dx^{i} dx^{j}\ .
\eeq
For convenience we turn to another coordinate system. Let us introduce $y=\tanh(r/\ell)$ and the EOM of $\gamma_{ij}$ becomes
\beq\label{eompp}
\left[\left( 1-y^{2}\right)\frac{d^{2}}{dy^{2}} -2y\frac{d}{dy} +\frac{m^{2}}{( -\Lambda )} +2-\frac{4}{\left( 1-y^{2}\right)}\right] \gamma_{ij} =0\ ,
\eeq
The mass of $4d$ graviton $m$ is defined by $4d$ wave equation
\beq
(\square _{4d}-2\Lambda )\gamma_{ij}=m^{2}\gamma_{ij}\ .
\eeq
After introducing another parameter $\nu=\left(\frac{9}{4} +\frac{m^2}{(-\Lambda)} \right)^{1/2} -\frac{1}{2}$, eq.~(\ref{eompp}) is simplified to
\beq
\left[\left( 1-y^{2}\right)\frac{d^{2}}{dy^{2}} -2y\frac{d}{dy} +\nu ( \nu +1) -\frac{4}{\left( 1-y^{2}\right)}\right] \gamma_{ij} =0\ .
\eeq
We can decompose $\gamma_{ij}$ into $\gamma_{ij} =\epsilon _{ij} f_m(y)$, then the solution of this equation of motion has a general form
\beq
f_m(y)=\alpha _{m} P_{\nu }^{2}( y) +\beta _{m} Q_{\nu }^{2}( y)\ ,
\eeq
where $P_{\nu}^{2}(y)$ and $Q_{\nu}^{2}(y)$ are the associated Legendre functions of the first and second kind. At the location of branes, we need to impose the Neumann boundary conditions,
\beq
\frac{1}{2}\partial_r\gamma_{ij}=\tanh(\frac{r}{l})\gamma_{ij}\ .
\eeq
Substitute the general solution at $r=0$ and $r=r_*$, we have
\beq
(\nu -1)(\nu +2)\left( \alpha _{m} P_{\nu }^{1}( 0) +\beta _{m} Q_{\nu }^{1}( 0)\right) =0\ ,
\eeq
\beq
y_*(\nu-1)\left(\alpha_{m}P_{\nu}^{2}(y_*)+\beta_{m}Q_{\nu}^{2}(y_*)\right)=(\nu-1)\left(\alpha_{m}P_{\nu+1}^{2}(y_*)+\beta_{m}Q_{\nu+1}^{2}(y_*)\right)\ ,
\eeq
where $y_*=\tanh(\frac{r_*}{\ell})$. These two boundary conditions will give us the mass spectrum, and it is obvious that the solution with $\nu=1$, which is $m=0$, satisfies these boundary conditions. The solution is
\beq\label{sol}
f_m(y(r))=\beta_{0}Q_{1}^{2}(y(r))=\beta_{0}\frac{2}{1-y(r)^2}=2\beta_{0}\cosh^2\left({r\over\ell}\right)\ ,
\eeq where $\beta_{0}$ is a finite constant. For $r\in(0,r_*)$, the zero mode is normalizable.

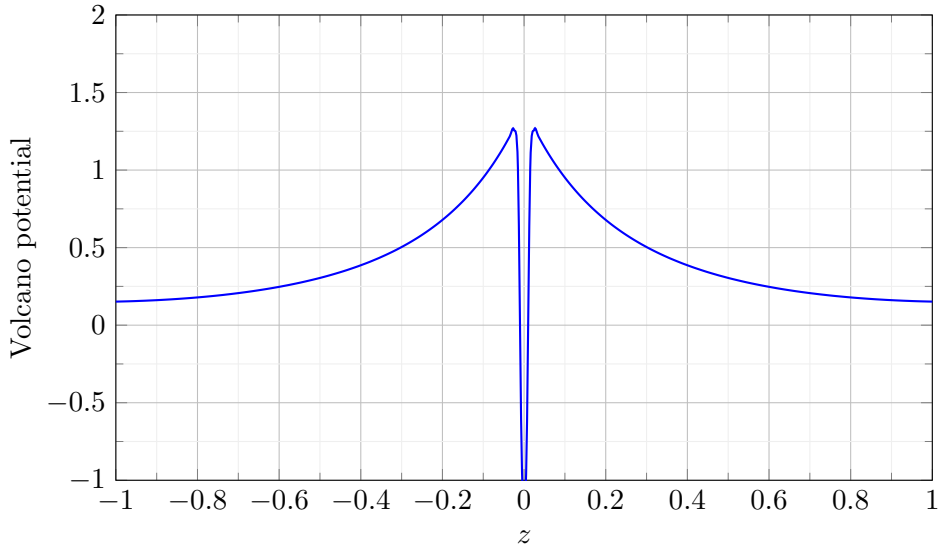
\begin{figure}
	\begin{center}
		\begin{tikzpicture}
			\begin{axis}[
				xmin = -1.0, xmax = 1.0,
				ymin = -1, ymax = 2.0,
				grid = both,
				minor tick num = 1,
				major grid style = {lightgray},
				minor grid style = {lightgray!25},
				width = 0.8\textwidth,
				height = 0.5\textwidth,
				xlabel = {$z$},
				ylabel = {Volcano potential},		]
				\addplot[
				domain = -1:1,
				restrict y to domain=-1:2,
				samples = 200,
				smooth,
				thick,
				blue,
				] {(15/40)*sin(deg(abs(x)+0.5))^(-2)-(3)*exp(-(abs(x)/0.01)^2)-(9/40)};
			\end{axis}
		\end{tikzpicture}
	\end{center}
	\caption{The volcano potential. The positive tension brane locates at $z=0$ and the reference brane locates at $z=1$. Note that there is a $\mathbb{Z}_2$ symmetry at $z=0$.} \label{M5}
\end{figure}

The EOM of $\gamma_{ij}$ can also be transformed into a  Schrodinger-like equation with volcano potential \cite{Karch:2000ct}. We need to rescale the perturbation $\hat{\gamma}_{ij}=\gamma_{ij}e^{-2A(r)}$ and the EOM of $ \hat{\gamma}_{ij} $ is
\beq\label{eomp}
\left( \partial _{r}^{2} +4A'\partial _{r}+e^{-2A}m^2\right) \hat{\gamma}_{ij} =0\ .
\eeq
Next we need to perform another coordinate transformation
\beq
z(r) =\frac{1}{\sqrt{-\Lambda }}\left\{\arcsin\left(\frac{1}{\cosh\left(\frac{r}{\ell}\right)}\right) -z_{0}\sqrt{-\Lambda }\right\}\ ,
\eeq
\beq
z_{0} =\frac{1}{\sqrt{-\Lambda }} \arcsin\left(\frac{1}{\cosh(r_*/\ell)}\right)
\eeq
and then use $H_{ij}(z) =e^{3A(z)/2} \hat{\gamma}_{ij}$, the EOM finally becomes
\beq
\left( -\partial _{z}^{2} +\frac{9}{4} A'(z)^{2} +\frac{3}{2} A''( z)\right) H_{ij}( z) =m^{2} H_{ij}(z)\ .
\eeq
The effective potential of this equation is
\beq
\bal
V(z)&=\frac{9}{4} A'( z)^{2} +\frac{3}{2} A''(z)\\
&=-\frac{9}{4}( -\Lambda ) +\frac{15}{4}\frac{( -\Lambda )}{\sin^{2}\left(\sqrt{-\Lambda }(z+z_{0})\right)} -3\cot\left(\sqrt{-\Lambda } z_{0}\right)\sqrt{-\Lambda } \delta (z)\ ,
\eal
\eeq
where $z\in \left( 0,\frac{\pi }{2\sqrt{-\Lambda }} -z_{0}\right)$ and we have used a delta potential. The positive tension brane locates at $z=0$ and the zero tension brane locates at $z=\frac{\pi }{2\sqrt{-\Lambda}}-z_{0}$. The potential between the positive tension brane and the zero tension brane is shown in Fig.~\ref{M5}. We can see that there is a potential trap at the positive tension brane, which can localize the massless graviton. The zero mode solution can be read from eq.~(\ref{eomp}) easily, which is a constant $\hat\gamma$. This is consistent with the solution eq.~(\ref{sol}).

\subsubsection{Boundary entropy in AdS/BCFT}
\begin{figure}
	\centering
	\includegraphics[scale=0.5]{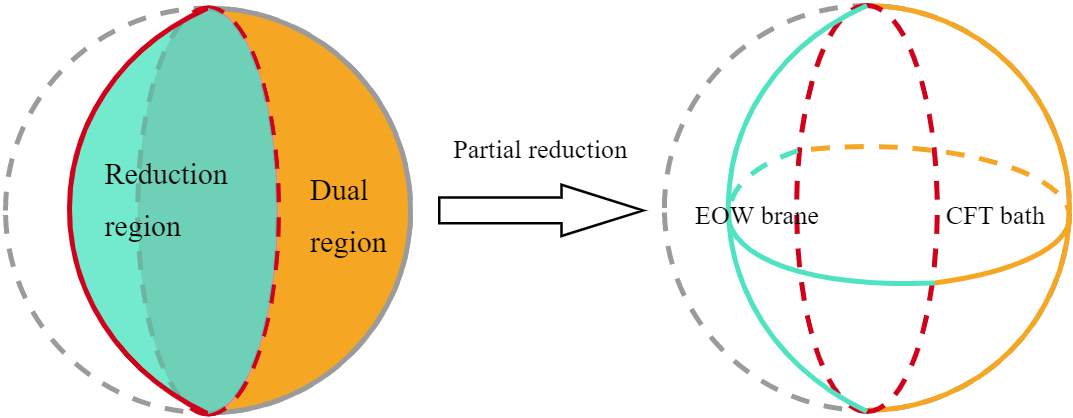}\\
	\caption{The partial reduction procedure of global AdS on a time slice. For AdS$_5$, a time slice of the BCFT$_4$ is defined on a $3d$ hemisphere with a S$^2$ boundary.}\label{pr}
\end{figure}

The partial reduction can be applied in AdS$_5$/BCFT$_4$ where the BCFT$_4$ is defined on half of the asymptotic boundary. The EOW brane can be introduced through Neumann boundary condition. This AdS$_4$ brane will intersect with the asymptotic boundary and give us a boundary CFT theory as showed in Fig.~\ref{pr}. In \cite{Kobayashi:2018lil}, the boundary entropy of this model has been calculated using on-shell action as well as RT surface. Now we review this process in the viewpoint of partial reduction and show a relation between boundary entropy and effective Newton constant.
In AdS$_5$/BCFT$_4$ duality, partial reduction implies that the holographic dual of the boundary defect is the reduction region, which is the spacetime between zero tension brane and EOW brane. So the information of the boundary defect, such as boundary entropy, can be recovered from the reduction region.
In Euclidean signature, the boundary entropy can be defined in terms of on-shell action
\beq\label{entropy}
S_{bdy}=-I_E(r_*)+\frac{1}{2}I_E(0)\ ,
\eeq
where the on-shell action is
\beq
I_E(r)=-\frac{1}{16\pi G^{(5)}_N}\int_{-\infty}^{r_*} d^{5} x\sqrt{g}( R-2\Lambda_5 ) -\frac{1}{8\pi G^{(5)}_N}\int_{Q} d^{4} y\sqrt{h}(K-(d-1)T)\ .
\eeq
One can also view eq.~(\ref{entropy}) as the on-shell action of the reduction region. The result has been calculated in \cite{Kobayashi:2018lil}
\beq
S_{bdy}=\frac{\ell^3}{4G^{(5)}_N}\mathrm{Vol}(\mathbb{H}^2)\int_{0}^{r_*}\cosh^2(r/\ell)dr\ .
\eeq

\begin{figure}
	\centering
	\includegraphics[scale=0.5]{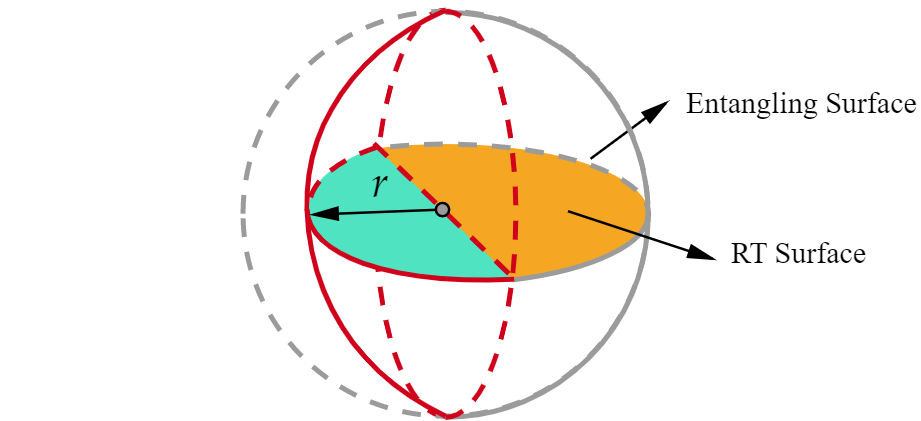}\\
	\caption{The colored region is the RT surface of $S^{\text{BCFT}}$. The RT of ${1\over 2}$$S^{\text{CFT}}$ corresponds to the orange region at the right side of dashed line.}\label{prt}
\end{figure}

We can also get this result by considering an entangling surface which splits the BCFT in half. Now we turn to Lorentzian signature with metric
\beq
ds^2=dr^2+\cosh^2\left({r\over\ell}\right)(-\cosh^{2}\rho dt^{2} +d\rho^{2} +\sinh^{2}\rho d{\Omega}_{2}^{2})\ ,
\eeq
and use
\beq\label{bentropy}
S_{bdy}=S^{\mathrm{BCFT}}-\frac{1}{2}S^{\mathrm{CFT}}\ ,
\eeq which is essentially the area of RT between the two branes as showed in Fig.~\ref{prt}. Note that the domain of $r$ is $(-\infty,r_*)$. So the boundary entropy is
\beq
\bal
S_{bdy}&=\frac{\ell^3}{4G^{(5)}_N}\mathrm{Vol}(\mathbb{H}^2)(\int_{-\infty}^{r_*}\cosh^2(r/\ell) dr-\int_{-\infty}^0\cosh^2(r/\ell) dr)\\
&=\frac{\ell^3}{4G^{(5)}_N}\mathrm{Vol}(\mathbb{H}^2)\int_{0}^{r_*}\cosh^2(r/\ell) dr\ ,
\eal
\eeq
in agreement with the on-shell action result.
Alternatively, the holographic dual of the boundary entropy can be viewed as an extremal surface between two branes. We consider an extremal surface attached freely to both the finite tension brane and the zero tension brane. Extremal condition demands orthogonal condition. This can be seen clearly in Fig.~\ref{prt}. We can obtain the effective Newton constant from this boundary entropy using Bekenstein-Hawking formula
\beq
\bal
G^{(4)}_{N}&=\frac{\mathcal{A}_4}{4S_{bdy}}
&=G^{(5)}_{N}\frac{4\cosh^2r_*/\ell}{2r_*+\ell \sinh(2r_*/\ell)}\ .
\eal
\eeq
which is the same as the result from eq.~(\ref{new}). Here $\mathcal{A}_4$ is the intersecting area of EOW brane and RT surface. Notice that the AdS radius on the brane is $\ell^2\cosh^2r_*/\ell$.
\subsection{Partial reduction and boundary entropy for cosmology}\label{sec3}

\begin{figure}
  \centering
  \includegraphics[scale=0.5]{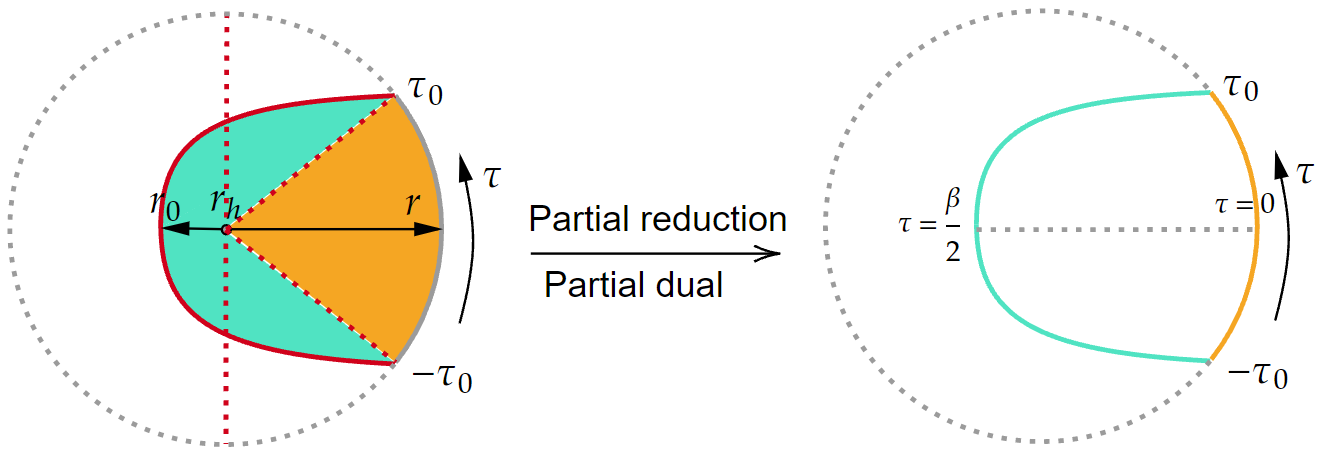}\\
  \caption{Partial reduction procedure for $d=4$ cosmology. The red solid line is a finite tension brane. The red dash line is the antipodal zero tension brane and the cusp zero tension brane. The orange region is dual to the boundary and the green region is reduced to the EOW brane.}\label{FIGPartialreduction4dBH}
\end{figure}

Now let us turn to the brane model for cosmology as the $4d$ generalization of Sec.~\ref{sec2}. For $d=4$, the bulk is an Euclidean AdS$_5$-Schwarzschild black hole,
\beq
	ds^{2} =f(r) d\tau ^{2} +\frac{dr^{2}}{f(r)} +r^{2}(d\theta^2+\sin^2\theta d\Omega_{2}^{2})
\eeq
with
\beq
	f(r)=\frac{r^2}{\ell^2}+1-\frac{r_h^{2}}{r^{2}}(\frac{r_h^2}{\ell^2}+1)\ .
\eeq
The periodicity of the $ \tau $ direction determines the temperature of the black hole, which is
\beq
	\beta =\frac{2\pi r_h \ell^2}{\ell^2+2r_h^2}\ .
\eeq
The brane trajectory $r(\tau)$ satisfies the Neumann boundary condition eq.~(\ref{beq}) as
\beq\label{beq4d}
	\frac{dr}{d\tau}=\pm\frac{r^4+r^2\ell^2-r_h^2(r_h^2+\ell^2)}
{T\ell^2r^3}\sqrt{\frac{r^4+r^2\ell^2-r_h^2(r_h^2+\ell^2)}{\ell^2r^2}-T^2r^2}\ .
\eeq
The brane trajectory is symmetric about $\tau=0$, where $r(\tau )$ takes its minimal value, which is determined by $f(r_0)=T^2r_0^2$ as
\beq\label{Tr0rh}
	r_{0} =\frac{1}{\sqrt{2(1-T^2\ell^2)}}\left[\sqrt{\ell^4+4(1-T^2\ell^2)r_h^2(\ell^2+r_h^2)}-\ell^2\right]^{\frac{1}{2}}\ .
\eeq
The Euclidean time $\tau _{0}$ can be obtained by subtracting half of the brane time from half of the whole time. It is related to the black hole temperature as
\beq
	\bal
		\tau _{0} & =\frac{\beta }{2} -\int _{r_{0}}^{\infty }\frac{Tr}{f(r)\sqrt{f(r)-T^{2}r^{2}}}\\
		& =\frac{\pi r_h \ell^2}{\ell^2+2r_h^2}-\int _{r_{0}}^{\infty }\frac{Tl^2r^3}{\left[r^4+r^2\ell^2-r_h^2(r_h^2+\ell^2)\right]
\sqrt{\frac{r^4+r^2\ell^2-r_h^2(r_h^2+\ell^2)}{\ell^2r^2}-T^2r^2}}\ .
	\eal
\eeq
Now for fixed $r_h$, the brane trajectory will intersect with the boundary of $(r,\tau)$ disk at different $\pm\tau_0$ for different tension $T$~\cite{Cooper:2018cmb}. A sketch of the brane trajectory is shown in Fig.~\ref{FIGPartialreduction4dBH}.

Using eq.~(\ref{beq4d}) the induced metric on the brane is
\beq
	\bal
        ds^{2} &=\left[f(r)+\frac{f(r)(f(r)-T^{2}r^{2})}{T^2r^2}\right]d\tau ^{2} +r^{2}(d\theta^2+\sin^2\theta d\Omega_{2}^{2})\\
        &=\frac{f(r)^2}{T^2r^2}d\tau^2+r^{2}(d\theta^2+\sin^2\theta d\Omega_{2}^{2})\\
        &=d\tau_p^2+r^2(\tau_p)(d\theta^2+\sin^2\theta d\Omega_{2}^{2})
	\eal
\eeq
with
\beq
r(\tau_p)=\frac{1}{\sqrt{2(1-T^2\ell^2)}}\left[\cosh(2\sqrt{1-T^2\ell^2}
\frac{\tau_p}{\ell})\sqrt{\ell^4+4(1-T^2\ell^2)r_h^2(\ell^2+r_h^2)}-\ell^2\right]^{\frac{1}{2}}\ .
\eeq
We can check that it is a hyperbolic space with scalar curvature $R_{brane} = -\frac{12(1-T^2\ell^2)}{\ell^2}$. This metric is conformally equivalent to the finite cylinder metric $ds_{brane}^{2}=\Omega^{-2} (\tau ')ds_{cyl}^{2} =\Omega^{-2} (\tau ')(\frac{1}{\ell^2}d\tau ^{\prime 2} +d\theta^2+\sin^2\theta d\Omega_{2}^{2})$ with a conformal factor $\Omega (\tau ')$ which we need to obtain by numerical.

Let us apply an analytical continuation to $\tau$. The time becomes $t=-i\tau$ and the brane world volume metric is
\beq
	\bal
        ds^{2} &=\left[-f(r)+\frac{f(r)(T^{2}r^{2}-f(r))}{T^2r^2}\right]dt ^{2} +r^{2}(d\theta^2+\sin^2\theta d\Omega_{2}^{2})\\
        &=-\frac{f(r)^2}{T^2r^2}dt^2+r^{2}(d\theta^2+\sin^2\theta d\Omega_{2}^{2})\\
        &=-dt_p^2+r^2(t_p)(d\theta^2+\sin^2\theta d\Omega_{2}^{2})\ ,\\
	\eal
\eeq
where
\beq
r(t_p)=\frac{1}{2\sqrt{(1-T^2\ell^2)}}\left[\cos(2\sqrt{1-T^2\ell^2}
\frac{t_p}{\ell})\sqrt{\ell^4+4(1-T^2\ell^2)r_h^2(\ell^2+r_h^2)}-\ell^2\right]^{\frac{1}{2}}
\eeq
is the scale factor which describes the $4$-dim big-bang big-crunch cosmology.

In the $d=4$ cosmology model, we can find that the equation for the zero tension brane in Euclidean $(r,\tau)$ plane is given by constant $\tau$. Inspired by the pure AdS case, we choose the $\tau=\pm\tau_0$ zero tension brane as the reference for the partial reduction. There is a cusp in the origin, but we can regularize this cusp in a $\epsilon$ small region near the origin. This can be justified by taking the $r\to r_h$ first in eq. (\ref{beq4d}).
The partial reduction procedure is shown in Fig.~\ref{FIGPartialreduction4dBH}.

Now we calculate the boundary entropy for the Euclidean cosmology boundary state and derive the brane Newton constant. In Euclidean $(r,\tau)$ plane, similar to the brane embedded in global AdS shown in Fig.~\ref{prt}, the boundary entropy can be dual to an extremal surface between two branes.
It can be verified that the extremal surface between the two branes is located at $\tau=0\ , \theta=\frac{\pi}{2}$. The area of this extremal surface is
\beq
    \bal
        S_{bdy}&=\frac{1}{4G_5}\mathrm{Vol}(\mathbb{S}^2)\int_{r_h}^{r_0}\frac{r^2}{f(r)^{\frac{1}{2}}}\\
        &=\frac{1}{4G_5}\mathrm{Vol}(\mathbb{S}^2)\int_{r_h}^{r_0}dr\frac{r^2}{\sqrt{\frac{r^2}{\ell^2}+1-\frac{r_h^{2}}{r^{2}}(\frac{r_h^2}{\ell^2}+1)}}\\
        &=\frac{1}{4G_5}\mathrm{Vol}(\mathbb{S}^2)\frac{1}{2}(Tr_0^2\ell^2-\ell^3\arsinh\frac{\sqrt{r_0^2-r_h^2}}{\sqrt{2r_h^2+\ell^2}})\ .
    \eal
\eeq
Treating this boundary entropy as an area entropy, we obtain the Newton constant given by
\beq
\frac{1}{4G_4}=\frac{S_{bdy}}{\mathrm{Vol}(\mathbb{S}^2)r_{0}^2}=\frac{1}{4G_5}\frac{1}{2}(T\ell^2-{\ell^3\over r_0^2}\arsinh\frac{\sqrt{r_0^2-r_h^2}}{\sqrt{2r_h^2+\ell^2}})\ .
\eeq Notice that there is a constrain among $T$, $r_0$ and $r_h$ in eq.(\ref{Tr0rh}).

\begin{figure}
  \centering
  \includegraphics[scale=0.35]{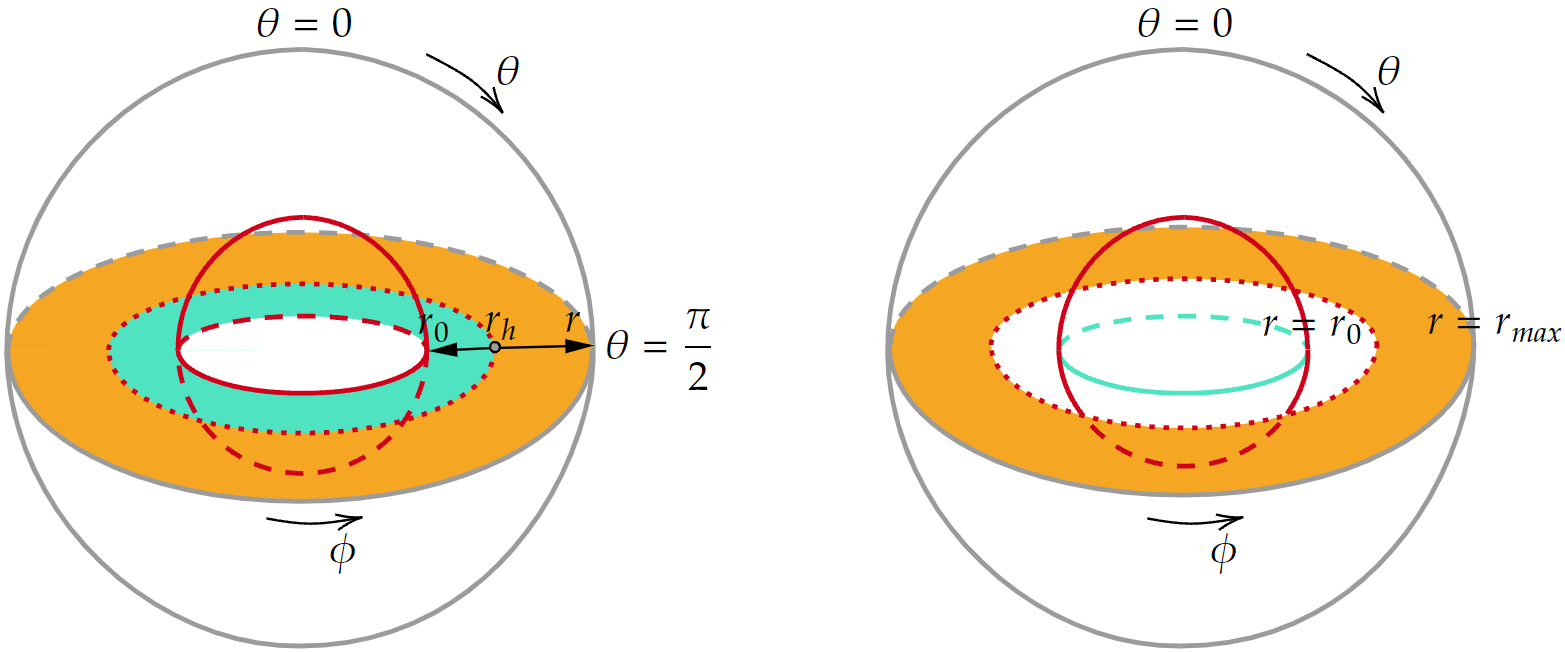}\\
  \caption{Left:~RT surface for entangling surface at $\theta=\frac{\pi}{2}$ in $t=0$ slice.\;Right:~after partial reduction, the boundary of island stay in $t=0$ slice with $\theta=\frac{\pi}{2}$.}\label{FIGRT4D}
\end{figure}

One can also justify the Newton constant in Lorentzian spacetime. Let us focus on the $\tau=0$ slice which is the same as $t=0$ and compute the entanglement entropy for the equal bipartition of $\theta$ direction from two different approaches.
Holographically this is given by the minimal surface between EOW brane and asymptotical boundary. On the other hand we can compute it using island formula after partial reduction. Ignoring the quantum matter located on the EOW brane,
the island result is given by the sum of the Area terms and minimal surfaces between zero tension brane and asymptotical boundary. Demanding the equivalence between the two approaches, one can obtain the same Newton constant as that in Euclidean case.
This is illustrated in Fig.~\ref{FIGRT4D}. It would be very interesting to study further whether the Newton constant has a time dependence. We leave this for future work.

\section{Conclusion and discussions}
We propose {\it partial reduction} as a new mechanism to localize gravity on the brane embedded in warped space. This was motivated by the recent development in information theoretical understanding of AdS$_{d+1}$/BCFT$_d$~\cite{Deng:2020ent,Chu:2021gdb,Li:2021dmf}. By turning on conformal matter on the brane, we have three different views of the same system: The first is AdS bulk gravity with a defect EOW brane (including the quantum matter on it); the second is the $d$-dimensional brane world gravity from partial reduction plus quantum matter on it, coupled to a CFT$_d$ bath coming from the dual of the remaining AdS region; the third is a complicated BCFT$_d$ with everything in the bulk encoded. These three different descriptions are equivalent. Partial reduction plays important role in clarifying the equivalence among the three different descriptions, in contrast with earlier attempts to understanding the boundary dual of Karch-Randall brane~\cite{Karch:2000gx,Aharony:2003qf}, as well as the recent double holography picture~\cite{Chen:2020uac,Chen:2020hmv,Geng:2020qvw,Geng:2020fxl}.

In this paper we focus on the application of {\it partial reduction} in cosmology. We first consider the cosmology from partial reduction in $d=2$. In particular we compute the fine grained entropy for a large subregion in CFT bath from two different approaches: One is the bulk defect extremal surface formula and the other is island formula. We compare those results and find exact agreement. We then move to $4d$ FRW cosmology where the brane trajectory is slightly complicated. We propose the proper partial reduction for the $4d$ cosmology and compute the Newton constant on the brane. We also show that partial reduction leads to a normalizable zero mode (massless graviton) on Karch-Randall brane in pure AdS$_5$ and suspect that the same thing will happen in $4d$ FRW cosmology.

There are a few future questions listed in order: First, extend our partial reduction mechanism to de-Sitter brane. In pure AdS$_5$, if one increases the brane tension beyond the critical value, the de-Sitter brane will appear~\cite{Karch:2020iit} and it would be interesting to study the localized gravity from partial reduction. Second, apply partial reduction to a pure Lorentzian cosmology including de-Sitter brane as well as a bath. This will tell us the fine grained entropy of a de-Sitter cosmology. Last but not least, following our construction from partial reduction, using defect and boundary CFT techniques one can try to bootstrap~\cite{Collier:2021ngi,Kusuki:2021gpt} the evolutional cosmology, which will eventually open the door to simulating our observational cosmology on quantum computer.

\section*{Acknowledgements}
We are grateful for useful discussions with our group members in Fudan University. This work is supported by NSFC grant 11905033. YZ is also supported by NSFC 12047502,11947301 through Peng Huanwu Center for Fundamental Theory.

\bibliographystyle{JHEP}
\providecommand{\href}[2]{#2}\begingroup\raggedright\endgroup

\end{document}